\documentclass[conference]{IEEEtran}
\IEEEoverridecommandlockouts
\usepackage{cite}
\usepackage{amsmath,amssymb,amsfonts}
\usepackage{algorithmic}
\usepackage{graphicx}
\usepackage{textcomp}
\usepackage{multirow}
\usepackage{array}
\usepackage{longtable}
\usepackage{bbding}
\usepackage{amssymb}
\usepackage{diagbox}
\usepackage{booktabs}
\usepackage[svgnames]{xcolor}
\usepackage{float}
\usepackage{url}
\def\BibTeX{{\rm B\kern-.05em{\sc i\kern-.025em b}\kern-.08em
    T\kern-.1667em\lower.7ex\hbox{E}\kern-.125emX}}

\begin{document}

\title{Smart Contract Vulnerability Detection Technique: A Survey}

\author{
    Peng Qian\IEEEauthorrefmark{1}\IEEEauthorrefmark{2}, 
    Zhenguang Liu\IEEEauthorrefmark{1}\IEEEauthorrefmark{2},
    Qinming He\IEEEauthorrefmark{1},
    Butian Huang\IEEEauthorrefmark{2},
    Duanzheng Tian\IEEEauthorrefmark{2},
    Xun Wang\IEEEauthorrefmark{2}\\
	\IEEEauthorrefmark{1}College of Computer Science and Technology, Zhejiang University, Hangzhou, China\\
	\IEEEauthorrefmark{2}School of Computer and Information Engineering, Zhejiang Gongshang University, Hangzhou, China\\
}

\maketitle

\begin{abstract}
Smart contract, one of the most successful applications of blockchain, is taking the world by storm, playing an essential role in the blockchain ecosystem. However, frequent smart contract security incidents not only result in tremendous economic losses but also destroy the blockchain-based credit system. The security and reliability of smart contracts thus gain extensive attention from researchers worldwide. In this survey, we first summarize the common types and typical cases of smart contract vulnerabilities from three levels, \emph{i.e., Solidity code layer, EVM execution layer, and Block dependency layer}. Further, we review the research progress of smart contract vulnerability detection and classify existing counterparts into five categories, \emph{i.e.,} formal verification, symbolic execution, fuzzing detection, intermediate representation, and deep learning. Empirically, we take 300 real-world smart contracts deployed on Ethereum as the test samples and compare the representative methods in terms of accuracy, F1-Score, and average detection time. Finally, we discuss the challenges in the field of smart contract vulnerability detection and combine with the deep learning technology to look forward to future research directions.
\end{abstract}


\section{Introduction}
\label{sec:introduction}
\IEEEPARstart{B}{lockchain} has become one of the most prominent technologies in the past few years, attracting worldwide attention~\cite{swan2015blockchain,zheng2018blockchain}. Essentially, a blockchain is a distributed shared transaction ledger, which is maintained by all the participant nodes in the blockchain network and is restricted by the consensus mechanism~\cite{sankar2017survey}. With the characteristics of decentralization, tamper-proof, and irreversibility, blockchain is being endowed the ability to reform the inherent mode of traditional industries, making breakthroughs in many fields, such as health care~\cite{xia2017bbds,azaria2016medrec,zhang2018blockchain}, copyright protection~\cite{meng2018design,holland2017copyright,qian2019digital}, supply chain~\cite{saberi2019blockchain,abeyratne2016blockchain,chen2017blockchain}, energy grid~\cite{mengelkamp2018blockchain,pop2018blockchain,knirsch2018privacy}, and Internet of Things~\cite{christidis2016blockchains,bahga2016blockchain,kshetri2017can}.

Smart contract~\cite{zheng2017overview,wang2019blockchain}, as one of the most successful applications of blockchain, has raised considerable enthusiasm in industry and academia. The concept of smart contracts can be traced back to 1994, which was first proposed by Nick Szabo~\cite{buterin2014next}. However, there is no available execution environment provided for smart contracts at that time until the emergence of blockchain technology. A smart contract is a computer protocol running on the blockchain, which is written by the \emph{Turing-complete} language, typically Solidity. So far, tens of thousands of smart contracts have been deployed on various blockchain platforms, \emph{e.g.,} Ethereum~\cite{wang2019blockchain}, EOS~\cite{jani2020smart}, VNT Chain~\cite{Etherscan}, and the number is still growing rapidly. Unfortunately, with the increasing number of smart contracts, security issues are undesirably emerging. On one hand, the code security problems of smart contracts may inevitably be introduced during code development. On the other hand, a smart contract deployed on the public blockchain is usually exposed in an open environment, making it easy to be attacked by hackers. Furthermore, due to the immutable and irreversible of smart contracts, we can only watch the funds flow into the attacker’s package and are unable to interrupt or prevent the contract execution when an attack occurs. 

According to the statistics from Bcsec~\cite{Bcsec} and Slowmist~\cite{Slowmist}, the economic losses caused by security issues in smart contracts have exceeded billions of dollars. For instance, in June 2016, hackers utilized the reentrancy vulnerability of the DAO (decentralized autonomous organization) contract~\cite{DAO} to steal around 60 million dollars worth of Ether (the digital currency of Ethereum). In July 2017, due to the {delegatecall} vulnerability of the Parity Multi-Sig Wallet contract~\cite{Parity}, Ether worth nearly 300 million dollars was frozen. In April 2018, malicious attackers exploited the integer overflow vulnerability of the Beauty Chain contract~\cite{BeautyChain} to issue an unlimited number of BEC tokens, leading to the evaporation of BEC value to zero. In May 2019, Binance Exchange~\cite{EOS} was compromised by hackers, resulting in the theft of more than 7,000 BTC. In recent months, smart contract games, \emph{e.g.,} FarmEOS, Playgames, LuckBet, EOSPlaystation~\cite{FarmEOS,Playgames,LuckBet,EOSlots}, have also suffered from attacks to a different extent, losing a total of nearly one million dollars. To summarize, the security problems of smart contracts not only lead to enormous financial losses but also destroy the fundamental trust based on blockchain applications. Therefore, effective security analysis and vulnerability detection methods for smart contracts are essential before their deployments. 

The reasons why smart contracts are particularly susceptible to attacks can be summarized in the following four aspects. \textbf{(1)} Current programming languages and tools for smart contracts (\emph{e.g.,} Solidity) are still nascent and primitive. Smart contracts written in such programming languages are relatively more difficult to check. Especially a smart contract is allowed to interact with external contract functions or interfaces, which may lead to repeated external invocations and unexpected security vulnerabilities. \textbf{(2)} Since smart contract developers cannot fully understand the basic execution logic of novel programming languages and audit tools, there may not be able to foresee all possible states and environments that a contract encounters in the future. This further makes the written contracts \emph{error-prone}. \textbf{(3)} In traditional programs, developers are able to modify bugs when they discover security problems. However, different from conventional programs, smart contracts and their states are stored in an immutable blockchain network so that smart contracts are unalterable. Once deployed, there is no way to update or modify the corresponding code of a smart contract unless manipulates more than 51\% computing power of the blockchain network (almost impossible). Therefore, this may make the contracts deployed on the blockchain network exist potential security issues but cannot be patched. \textbf{(4)} Since smart contracts host digital assets (such as Bitcoin~\cite{nakamoto2008bitcoin} and Ether) worth millions of dollars, which makes them attractive to many malicious attackers. High profits drive attackers to exploit the vulnerabilities in smart contracts to steal money. In summary, compared to other software or application, smart contracts are susceptible to security vulnerabilities and malicious attacks, which further highlights the importance of smart contract security protection.

\begin{figure}
	\centering
	\includegraphics[width=8.8cm]{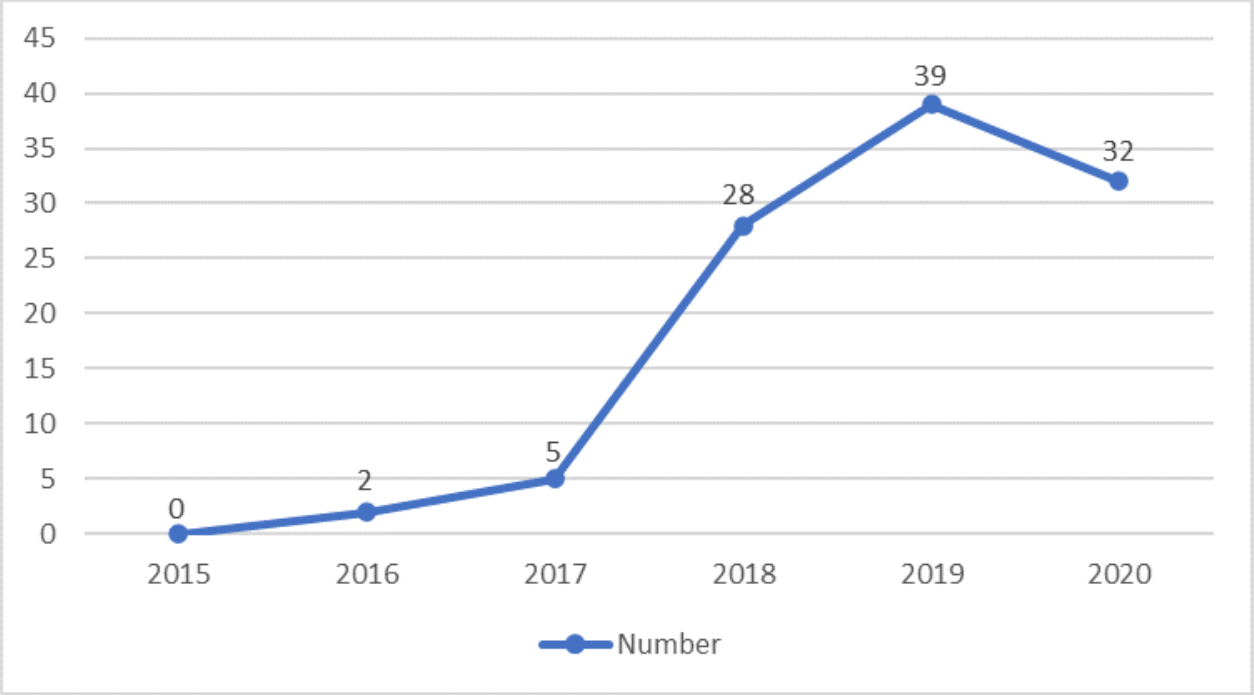}
	\caption{The tendency change of literature in the area of smart contract security analysis and vulnerability identification}
	\label{fig_Barchart}
\end{figure}

In this survey\footnote{This manuscript is the English translation version of our paper published in \emph{Ruan Jian Xue Bao/Journal of Software, 22, 33(8)}.}, we extensively investigate relevant works that propose a new method or theory in the area of smart contract security analysis and bug detection. These researches are summarized from authoritative databases, such as Web of Science, IEEE Xplore, and SpringerLink. Up to September 2020, there are more than one hundred relevant pieces of literature. It can be observed from Fig.~\ref{fig_Barchart} that the number of related works increased rapidly in the past three years.

The remainder of this survey is organized as follows. Section~\ref{sec:types and cases} classifies the smart contract vulnerabilities into three levels and restores five typical cases. Section~\ref{sec:analysis_method} reviews the research progress of smart contract vulnerability detection from five aspects, \emph{viz.} formal verification, symbolic execution, fuzzing detection, intermediate representation, and deep learning. Section~\ref{sec:tools} compares and analyzes the automation level, open-source nature, and performance (\emph{e.g.,} accuracy, F1-Score, average detection time) of various methods. Section~\ref{sec:conclusion} discusses the shortcomings of existing smart contract bug detection methods, and looks forward to future research challenges and directions in smart contract vulnerability identification.

\section{Smart Contract Bug Classes and Cases}
\label{sec:types and cases}
With the popularity of smart contracts and related decentralized applications (DAPP)~\cite{mohanta2018overview,zheng2017overview,macrinici2018smart,20191681}, the derived security vulnerabilities of smart contracts are being discovered and exploited by more and more malicious attackers~\cite{fu2019research,0Security,20223059,2020Smart}. Compared with traditional software programs, the security problems caused by smart contracts are more complicated, and the analysis and verification of smart contract vulnerabilities tend to be more difficult.

Ethereum~\cite{wang2019blockchain} is the most popular and influential open source public blockchain platform, thanks to it holding the largest number of smart contracts and excellent DAPPs. Dune Analytics~\cite{Dune} reported that there are more than two million smart contracts deployed on the Ethereum platform in March 2020. Moreover, according to the proportion statistics of the DAPP market in the first two quarters of 2020, Ethereum DAPP accounted for 82\% of the total created value, of which 80\% belong to high-risk industries such as gambling and games. In particular, Ethereum smart contracts present an extraordinary variety of security vulnerabilities, which have caused tremendous economic losses of more than one billion dollars. More importantly, most of the typical vulnerability events (\emph{e.g.,} The DAO) can be traced from Ethereum. Smart contracts of other platforms usually take Ethereum as the exemplar. In practice, developers and researchers regard Ethereum smart contracts as their primary research targets for security analysis and program verification. Therefore, we select Ethereum smart contracts as an example to analyze the specific security vulnerabilities and cases.

Technically, Ethereum smart contract vulnerabilities can be divided into three levels, \emph{i.e.,} Solidity code layer, EVM execution layer, and Block dependency layer~\cite{Hard-Fork}. In this section, we introduce fifteen kinds of Ethereum smart contract vulnerabilities, while the specific vulnerability types and illustrations are listed in Table~\ref{table_smart contract vulnerability}.

\renewcommand{\arraystretch}{1.05}
\begin{table*}
	\caption{Smart Contract Vulnerability Classification and Explanation. ‘—’ denotes not applicable}
	\centering
	\resizebox{0.955\textwidth}{!}{
		\begin{tabular}[htbp]{|m{2.8cm}<{\centering}|m{2.8cm}<{\centering}|m{4cm}<{\centering} |m{2.7cm}<{\centering}|m{3cm}<{\centering}|}
			\hline
			\textbf{Vulnerability Level}&\textbf{Vulnerability Type}&\textbf{Vulnerability Definition}&\textbf{Related Attack}&\textbf{Security Issue}\\
			\hline
			\multirow{15}{*}{Solidity code layer}&
			Reentrancy&The \emph{fallback} function has recursive calls to external contracts&The DAO Attack&Unable to store and protect contract tokens\\  
			\cline{2-5}  
			&Integer Overflow/Underflow &Value is out of the defined integer type range&Beauty Chain Integer Overflow Attack&Integer range error\\
			\cline{2-5}
			&Access Control&Function or variable access is restricted to public type&—&Arbitrarily calls to function or variable\\
			\cline{2-5}
			&Mishandled Exception&Return value and type are not checked after function call&The DAO Attack, KoET Attack&Exception handling failed\\
			\cline{2-5}
			&Denial of Service&Unexpected revert; \emph{Gas} exceeds upper limit; Unprotected \texttt{owner} account&KoET attack&Token frozen; unable to store and protect contract tokens\\
			\cline{2-5}
			&Type Mismatch&Variable type definition error&— &Unable to store and protect contract tokens\\
			\cline{2-5}
			&Unknown Function Call&The \emph{fallback} function is triggered by calling unknown functions or transferring money&The DAO Attack&Unable to store and protect contract tokens\\
			\cline{2-5}
			&Ether Frozen&Unauthorized use of contract self destruction&Parity Multi-Sig Wallet Attack&Inappropriate contract or function access\\
			\hline
			\multirow{6}{*}{EVM execution layer}&Short Address&Contract address fails to satisfy requirement (length is less than 20 digits)&—&Unable to store and protect contract tokens\\ 
			\cline{2-5}
			&Ether Loss&Wrong or empty contract address&—&Unable to store and protect contract tokens\\
			\cline{2-5}
			&Call-Stack Overflow&Exceed the upper limit of contract invocations&—&Buffer overflow\\
			\cline{2-5}
			&Transaction Origin Use&Use \texttt{tx.origin} for smart contract authentication&—&Inappropriate contract or function access\\
			\hline
			\multirow{4}{*}{Block dependency layer}&Timestamp Dependency&Assign block timestamp to variables&—&Fail to use secure random numbers\\  
			\cline{2-5}
			&Block Dependency&Assign block-related parameters to variables&—&Fail to use secure random numbers\\
			\cline{2-5}
			&Transaction Order Dependency&Inconsistent transaction sequence&—&Race conditions\\
			\hline
	\end{tabular}}
\label{table_smart contract vulnerability}
\end{table*}

\subsection{Solidity Code Layer}
\emph{(1) Reentrancy.}
Generally speaking, due to the atomicity and sequentiality of program execution, a new command will not be performed until the end of the current process. However, smart contracts do not comply with this rule. Malicious attackers are able to re-enter the called function during the current program execution~\cite{grossman2017online}. Similar to most programming languages, smart contracts engage in cross-function or cross-contract invocations to process business logic. But the difference is that smart contracts usually involve sensitive operations, \emph{e.g.,} money deposit or transfer. 

Furthermore, due to the default settings of smart contracts, the transfer operation will inevitably trigger the \emph{fallback} mechanism in the \emph{recipient} contract. When a smart contract performs a cross-contract operation of transferring money, attackers may capture such external invocation and perform some malicious operations. For example, an attacker designs malicious code in its \emph{fallback} function, which can recursively re-enter a \emph{victim} contract to call the \emph{transfer} function to steal Ether. We consider that the reentrancy vulnerability is considered as an external transfer invocation that can call back to itself through a chain of calls. Reentrancy vulnerability has resulted in the notorious security incident in the history of smart contracts (\emph{i.e.,} The DAO Attack~\cite{DAO}), which not only led to the losses of nearly 60 million dollars but also caused the \emph{hard-fork}~\cite{Hard-Fork} of Ethereum.

\emph{(2) Integer Overflow/Underflow.}
Integer overflow is a common vulnerability in many programs, usually divided into overflow and underflow. There are three types of integer overflow vulnerabilities in smart contracts, \emph{i.e.,} multiplication overflow, addition overflow, and subtraction underflow. In the source code of smart contracts, integers are treated as fixed-size and unsigned integer types, and the value of integer variables is limited to the range of the defined type. Obviously, if an integer variable exceeds a certain range, an integer overflow error will occur. 

Ethereum smart contracts are written in high-level languages, such as \emph{Solidity}, which supports the integer range from \emph{uint8} to \emph{uint256}. For example, if a number $v$ is of type \emph{uint8}, its value is stored as a 8-bits unsigned number ranging from 0 to $2^{8}-1$. If the value out of this range is assigned to a variable of \emph{uint8} type, Ethereum virtual machine (EVM) will automatically truncate the high digits. Different from other programs, the losses caused by integer overflow vulnerability in smart contracts are enormous and irreparable. For example, the integer overflow vulnerability was exploited by attackers who copy BEC tokens indefinitely, leading to the value of BEC evaporating to zero \cite{BeautyChain}. Currently, to prevent the integer overflow of smart contracts, developers are not only required to check the code manually but also employ the \emph{SafeMath} \cite{SafeMath} library to verify the arithmetic logic.

\emph{(3) Access Control.}
The fundamental reason for the access control vulnerability is that the access permissions of functions in the contract are not clearly or carefully checked, thereby allowing malicious attackers to utilize functions or variables that should not be accessed by them. Access control vulnerability is usually reflected in two levels: 1) Code Level. There are four types of access restrictions in smart contract functions and variables, namely \emph{public, private, external,} and \emph{internal}; 2) Logic Level. A modifier is usually used to constrain the access rights of functions in smart contracts, such as \emph{onlyOwner} and \emph{onlyAdmin}. Functions without modifier restrictions show that anyone has the right to access and manipulate them, which may cause the key functions to be manipulated by malicious attackers, thus endangering the security of smart contracts.

\emph{(4) Mishandled Exception.}
 Exception handling is one of the most important mechanisms to improve program reliability and robustness. In smart contracts, there are three kinds of exceptions. 1) \emph{Gas Exhausted}. Here, \emph{Gas} means the cost of deploying or executing a smart contract. When the \emph{Gas} was used up, an exception will be thrown. 2) \emph{Call-Stack Overflow}. When the number of invocations exceeds the maximum settings of EVM, a call-stack overflow exception will be triggered. 3) \emph{Exception Statement}. There is an exception handling instruction, \emph{e.g.,} \emph{throw} in the execution statement.
 
In general, a smart contract handles the abnormal behavior through rolling back, namely terminates the current contract execution, restores to the previous state, and returns an error identifier, \emph{i.e., false}. However, since there is no unified method for handling exceptions, the \emph{caller} contract may not be able to obtain the exception information from the \emph{callee} contract. For example, when an exception occurs in a sub-call of smart contracts, it will be propagated to its superior automatically. Nevertheless, some underlying function invocations (e.g., \texttt{send}, \texttt{delegatecall}) only return \emph{false} without throwing an exception. Therefore, it is not safe to judge whether the contract is successfully executed based on if the exception is thrown. We consider that the return value should be checked strictly when calling the underlying functions, and the exception should be handled consistently in the meantime.

\emph{(5) Denial of Service.}
Denial of service (DoS) \cite{praitheeshan2019security} is a common vulnerability of Ethereum smart contracts. Attackers usually exploit such vulnerability to destroy the original logic and consume additional resources, e.g., \emph{Gas} and \emph{Ether}, making the contract unable to provide normal services for a while or forever. There are usually three types of DoS attacks against smart contracts.
\begin{itemize}
\item DoS attack launched through (unexpected) revert. When the state update of the smart contract depends on the execution results of an external function, the smart contract will be susceptible to the DoS attack once the failure of external function execution is not handled in time.
\item DoS attack launched through the upper limit of gas. Each block in Ethereum has the upper limit of gas. A transaction initiated by the contract will be blocked as long as the cost of gas exceeds this limit. Therefore, even if there is no malicious attack, a smart contract may also have problems due to being out of the gas limit. More seriously, if an attacker maliciously manipulates the cost of gas so that the gas reaches the limit unintentionally, the transaction process of the contract will end in failure. 
\item DoS attack launched through the \texttt{owner} account of contracts. Most smart contracts have an \texttt{owner} account, which has the ability to control the contract. If the \texttt{owner} account is not protected, it is probably manipulated by attackers, which may further make the contract in danger (\emph{e.g.,} Ether is frozen permanently).
\end{itemize}

\emph{(6) Type Mismatch.}
Solidity is a strongly typed programming language, which can automatically check whether there is a type mismatch in the program. For example, if the value of a \emph{string} type is assigned to an \emph{integer} variable, a type mismatch will occur. In smart contracts, however, even if the type mismatch in some cases, the contract cannot raise an exception during program execution. Generally, developers default that the contract has already checked whether the type is matched in the program, thereby ignoring the manual audit and leading to unexpected vulnerability.

\emph{(7) Unknown Function Call.}
Similar to most programming languages, a smart contract ensures the uniqueness of functions by matching the function name and parameter number. When a contract calls a function in an external contract, if the function name and parameter number fail to match any function in the callee contract, the fallback function in this contract will be triggered automatically~\cite{praitheeshan2019security}. At this time, a security problem may occur unexpectedly, if the malicious operation designed by the attacker is hidden in the fallback function.

\emph{(8) Ether Frozen.}
Transfer operation is one of the important and unique abilities of smart contracts which means the smart contract can receive or transfer Ether from/to other contracts. It is worth mentioning that some contracts do not need to implement the \emph{transfer} function themselves. Instead, they can utilize \texttt{delegatecall} to call a transfer function in an external contract to achieve the transfer operation. However, if the called external transfer function has self-destruction operations such as \emph{Self-destruct} and \emph{Suicide}, Ether on the current contract (which calls this \emph{transfer} function through \texttt{delegatecall}) is likely to be frozen due to such a self-destruction operation executed.

\begin{figure}
	\centering
	\includegraphics[width=8.8cm]{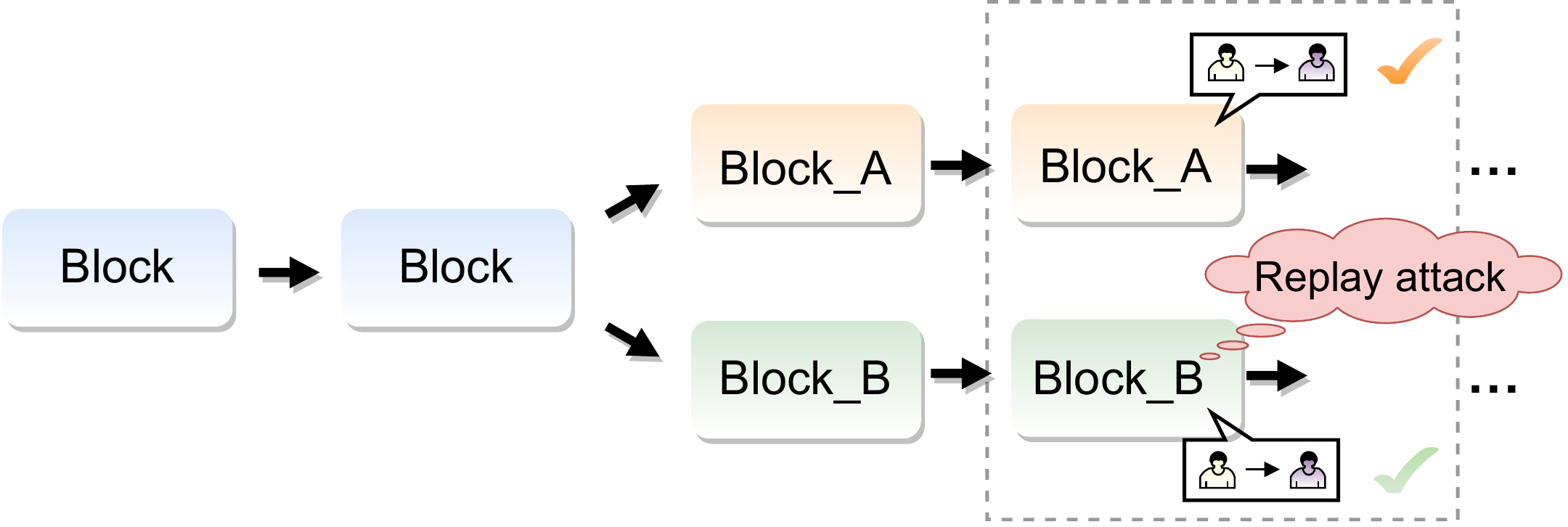}
	\caption{A simplified illustration of the replay attack}
	\label{fig_replay}
\end{figure}

\emph{(9) Replay Attack.}  
Replay attack~\cite{ReplayAttack} was discovered after the Ethereum \emph{hard-fork}. Since the address, private key, algorithm, and transaction format are the same before the \emph{hard-fork}, transactions initiated on another chain are also valid. As shown in Fig.~\ref{fig_replay}, a user who received a certain number of tokens from someone else through one ledger could switch to another ledger, which causes the transaction replicated, and an identical amount of tokens will be fraudulently transferred to the user's account again.

\subsection{EVM Execution Layer}
\emph{(1) Short Address.}
The length of a contract address is 20 digits, which is defined by the Ethereum ABI specification~\cite{atzei2017survey}. When the length of a smart contract address is less than a regular one, EVM will fill zero automatically so that the length of the address is equal to 20. This may leave a large room for malicious attackers. 
For example, the attacker deliberately inputs a short address, which triggers the EVM to take the missing coding digits from the next parameter (\emph{i.e., Ether amount}) to complete the address, and then fill the end of the whole string of binary codes with 0. This indicates that the parameter of \emph{Ether amount} has been shifted to the left by one byte. If the transfer operation is performed at this time, the contract may be transferred out to the attacker more than the actual Ether that should be forwarded.

\emph{(2) Ether Loss.}
When a smart contract transfers Ether, the recipient contract address needs to be specified and standardized. Assuming this address is an independent empty address and is not associated with any other users or contracts, it probably causes the Ether to lose forever once Ether is transferred to such a contract address.

\emph{(3) Call-Stack Overflow.}
When a smart contract calls an external function or executes a self-call, this will increase the call-stack depth of the contract. In the Ethereum virtual machine, the upper limit of the call stack is 1,024. Therefore, if an attacker designs a series of nested calls, it probably causes the call-stack overflow, which further leads to unforeseen security problems.

\begin{figure}
	\centering
	\includegraphics[width=8.7cm]{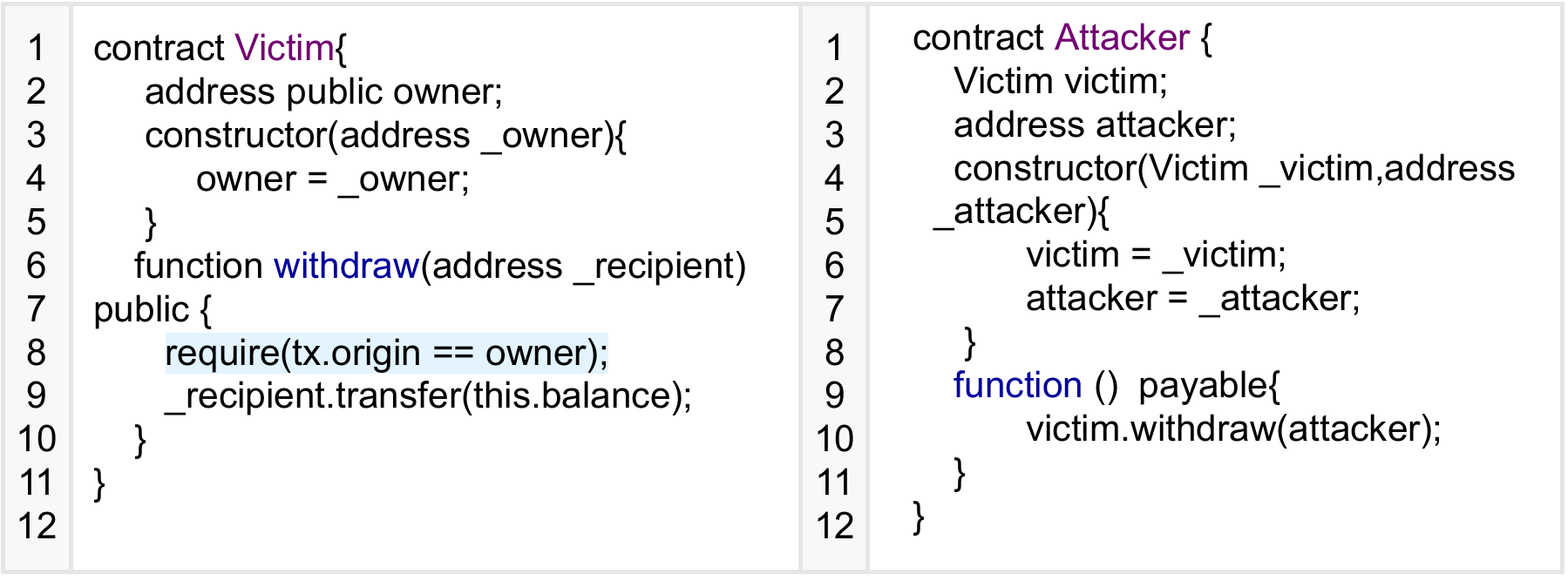}
	\caption{A simplified example of the \texttt{tx.origin} vulnerability}
	\label{fig_txorigin}
\end{figure}

\emph{(4) Transaction Origin Use.}
Ethereum smart contract has a global variable, i.e., \texttt{tx.origin}, which can backtrack the entire call stack and return the original address of the called contract. If a contract uses this global variable for authorization and authentication, attackers can exploit the characteristics of \texttt{tx.origin} to design the corresponding attack to steal Ether. We describe a specific case of \texttt{tx.origin} vulnerability shown in Fig.~\ref{fig_txorigin}. An attacker calls the \emph{withdraw} function of the \emph{Victim} contract in its \emph{fallback} function to induce the \emph{Victim} contract to transfer Ether to the \emph{Attacker} contract. However, due to the statement of ``{\small$tx.origin==owner$}'' (line 8 in the left of Fig.~\ref{fig_txorigin}), it is difficult to detect anomalies so that all the Ether in the \emph{Victim} contract are transferred to the \emph{Attacker}.

\begin{figure}
	\centering
	\includegraphics[width=8.7cm]{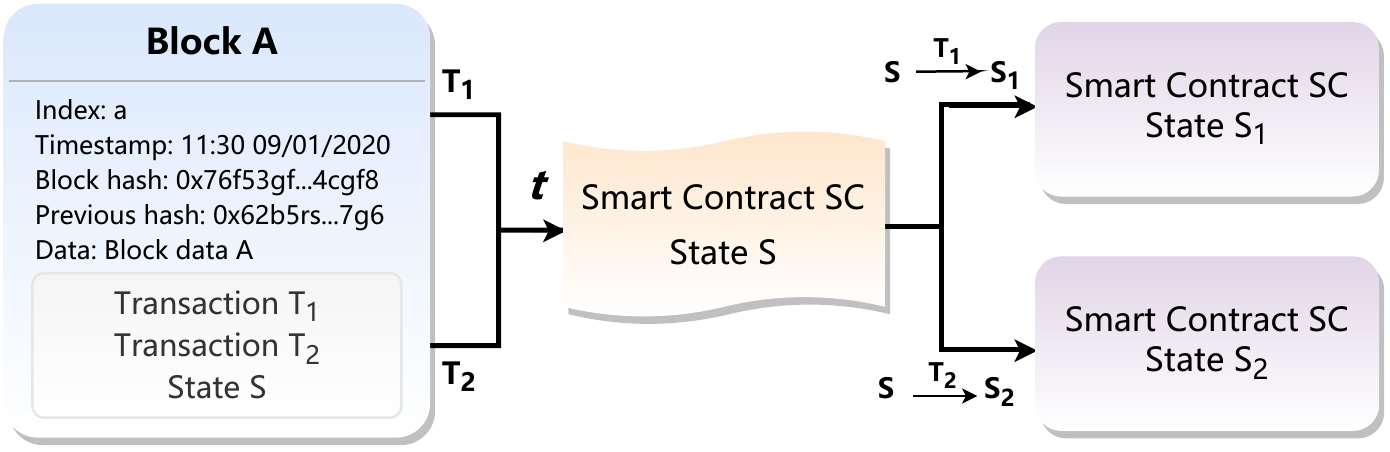}
	\caption{An example of the transaction ordering dependency}
	\label{fig_TOD}
\end{figure}

\subsection{Block Dependency Layer}
\emph{(1) Timestamp Dependency.}
 Smart contracts usually utilize the block timestamp confirmed by miners (\emph{i.e.,} nodes in the blockchain network) to achieve time constraints. All transactions in the block share the same timestamp, which ensures the consistency of the contract state. However, miners who confirm the block may deliberately choose a timestamp that is beneficial for themselves to grab benefits.

\emph{(2) Block Dependency.}
Ethereum smart contracts cannot directly call the built-in functions in smart contracts to generate a random number. Therefore, developers tend to use the block parameters, such as block number (\emph{block.number}), block timestamp (\emph{block.timestamp}), block hash (\emph{block.blockhash}), or other related block parameters as the basic seeds to implement a random number generation function. However, similar to the timestamp dependency, the block parameters can be manipulated in advance by attackers, resulting in the generated random number being predictable, which may be exploited by malicious attackers to produce random numbers beneficial to themselves.

\emph{(3) Transaction Order Dependency.}
We have already mentioned in the previous paragraph that miners can determine the order of transaction execution in the blockchain network. However, the state updating of smart contracts strictly depends on the order of transaction execution, and the wrong order may incur a negative impact. Here, we describe a specific case of transaction order dependency shown in Fig.~\ref{fig_TOD}. Users $A$ and $B$ respectively submit transactions $T_{1}$ and $T_{2}$ at time \textit{\textbf{t}}. Due to the execution order of $T_{1}$ and $T_{2}$ determined by miners, if $T_{1}$ is executed first, the contract state will be updated from $S$ to $S_{1}$ and vice versa. Therefore, the final contract state depends on the transaction execution order confirmed by miners. At this point, if a malicious miner monitors the contract transaction in the block, the updating of the current contract state is likely to be controlled by blocking corresponding transactions, thereby deploying an attack in advance.

\subsection{Smart Contract Security Incidents}
In the real world, there exist many well-known smart contract security incidents. Here, we elaborate on several typical cases that have caused tremendous economic losses and seriously hindered the development of smart contracts.

\begin{figure}
	\centering
	\includegraphics[width=8.75cm]{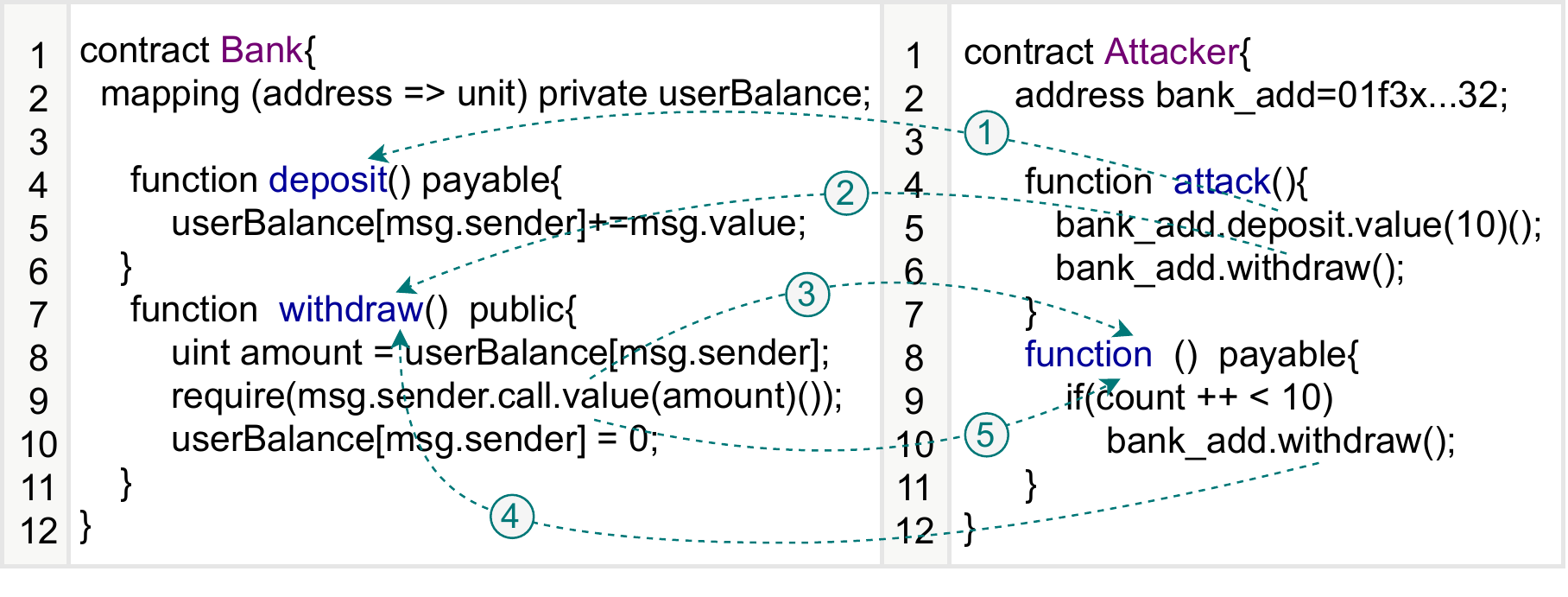}
	\caption{A simplified version of The DAO attack}
	\label{fig_theDao}
\end{figure}

\emph{(1) The DAO Attack.}
DAO is actually a decentralized autonomous organization that implements the DAO contract used for crowdfunding, which has already raised approximately 245 million US dollars of Ether before being attacked. In 2016, attackers exploited the reentrancy vulnerability of the DAO contract to steal the Ether worth 60 million US dollars, directly causing the \emph{hard-fork}~\cite{DAO} of Ethereum. In order to have a deeper understanding of the DAO attack, we present a simplified version as shown in Fig.~\ref{fig_theDao}.
Specifically, the \emph{Bank} contract has a \emph{userBalance} variable and two functions: {\emph{deposit}} and {\emph{withdraw}}. Function {\emph{deposit}} allows users to deposit money, while function {\emph{withdraw}} enables users to withdraw money via invoking {\emph{call.value}} (\emph{Bank}, line 9). To attack this contract, the \emph{Attacker} contract, shown on the right of Fig.~\ref{fig_theDao}, implements two functions, \emph{i.e.,} {\emph{attack}} and anonymous {\emph{fallback}} functions (\emph{Attacker}, lines 8--11). We would like to highlight that the {\emph{fallback}} function of the smart contract, namely the function without name and argument, will be automatically invoked when the contract receives any Ether.

\textbf{Attack.} As shown in Fig.~\ref{fig_theDao}, contract \emph{Attacker}  can steal Ether from contract \emph{Bank} through the following steps. First,  \emph{Attacker} deposits 10 Ether in contract \emph{Bank} by calling the \emph{deposit} function. Then, \emph{Attacker} withdraws the 10 Ether by invoking the \emph{withdraw} function (step 2). When the contract \emph{Bank} sends Ether to \emph{Attacker} using \emph{call.value} (\emph{Bank}, line 9), the \emph{fallback} function of \emph{Attacker} will be automatically invoked (step 3). In its \emph{fallback} function, \emph{Attacker} calls \emph{withdraw} again (step 4). Since the {\emph{userBalance}} of \emph{Attacker}  has not yet been updated (\emph{Bank}, line 10), \emph{Bank} believes that \emph{Attacker} still has Ether balance in the contract, thus transferring 10 Ether to \emph{Attacker} again (Step 5). The withdraw loop lasts for 9 times ({\small$count++ < 10$}, \emph{Attacker} line 11).  Finally, \emph{Attacker} obtains much more Ether (100 Ether) than expected (10 Ether).

\textbf{Underlying issue.} The underlying security problem is that the balance of \emph{Attacker} is updated after money transfer. Therefore, \emph{Attacker} can call {\emph{withdraw}} again utilizing the \emph{fallback} mechanism, making  contract \emph{Bank}  wrongly believe that it still has enough balance and transfer money to \emph{Attacker} again.

\emph{(2) King of the Ether Throne.}
King of the Ether Throne (KoET) is an Ethereum gambling game~\cite{KoET}, in which players can compete for the \emph{Ethereum throne} to win all the bonuses in the contract. Whenever a player sends a \emph{competition fee} (i.e., Ether) to compete for the \emph{throne}, the \emph{fallback} function in the KoET contract will be triggered. Then, the \emph{fallback} function will check whether the \emph{competition fee} is sufficient. If not, the contract throws an exception and rolls back this transaction. Otherwise, the player will become the new \emph{Ethereum throne}. In the meantime, the KoET contract needs to send part of the \emph{competition fee} as a reward to the former \emph{throne}, and the rest remains in the contract as the final bonus. After 24 hours, the player who competes successfully will receive all the bonuses in the KoET contract.

\begin{figure}
	\centering
	\includegraphics[width=8.75cm]{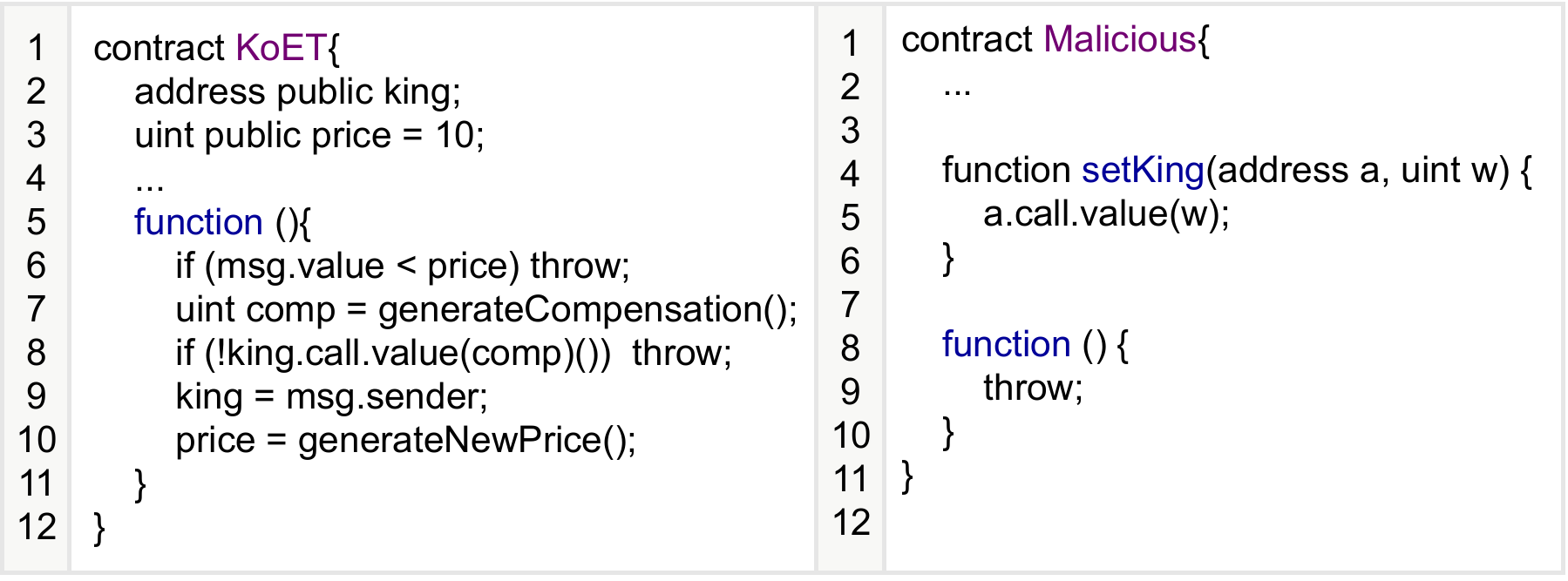}
	\caption{An example of King of the Ether Throne Attack}
	\label{fig_KoET}
\end{figure}

However, this gambling game is not as reasonable as it seems, which is susceptible to denial of service (DoS) attack. As shown in Fig.~\ref{fig_KoET}, the attacker will constantly occupy the \emph{throne} until he wins all the bonuses in the contract through the following three steps. 
\begin{itemize}
\item First, the attacker executes the \emph{setKing} function in the \emph{Malicious} contract to send enough \emph{competition fee} to compete for becoming a new \emph{throne}. 
\item Then, when a new player participates in the competition and pays a sufficient \emph{competition fee}, KoET contract will send rewards to the current \emph{throne} (i.e., \emph{Malicious}), further triggering the \emph{fallback} function in the \emph{Malicious}, where only exists a statement of exception throw. 
\item Finally, due to the exception handling, KoET tries to send rewards to \emph{Malicious} again. In the end, sending rewards multiple times will exhaust \emph{Gas}, and nobody can compete for the \emph{throne} again so the attacker occupies the \emph{throne} all the time and wins all the final bonuses.
\end{itemize}

\emph{(3) Parity Multi-Sig Wallet.}
The parity multi-signature wallet is a public library contract implemented to manage the digital asset of users, which includes some common functions and logic code. Users can call functions in this public library contract in their wallet contract to execute the corresponding business logic. However, centralized management of functions in the public library contract can easily become the target of attackers. Since most of the user wallets in Ethereum rely on the parity multi-signature wallet, attackers can disturb all the wallet contracts by attacking the public library contract.

\begin{figure}
	\centering
	\includegraphics[width=8.7cm]{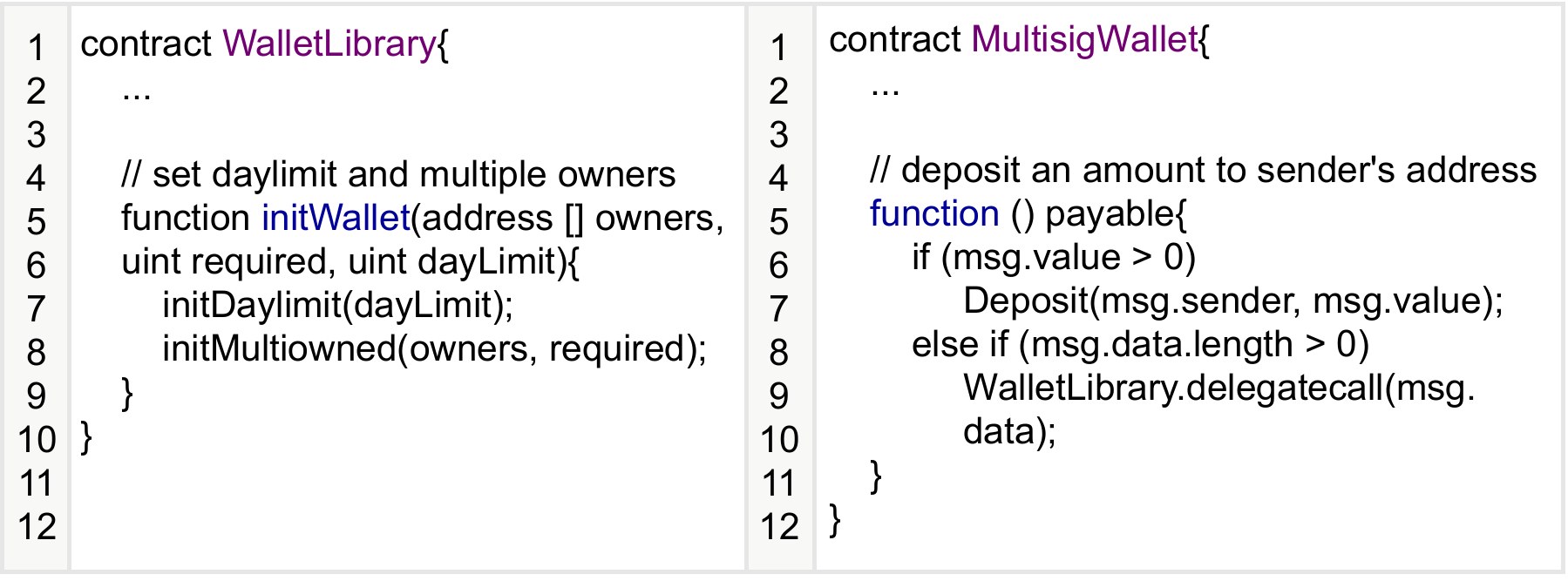}
	\caption{An example of Parity Multi-Sig Wallet Attack}
	\label{fig_Parity}
\end{figure}

Fig.~\ref{fig_Parity} describes the example of Parity Multi-Sig Wallet, including the contract fragment of \emph{WalletLibrary} and \emph{MultisigWallet}. In the \emph{WalletLibrary}, the \emph{initWallet} function is used to initialize the usage date and owner of the wallet. Other wallet contracts can use \emph{delegatecall} to call the \emph{initWallet} function. As shown in line 9 of the \emph{MultisigWallet} contract, the wallet uses \emph{delegatecall} to call the public library \emph{WalletLibrary}. Since all the public functions in \emph{WalletLibrary}, such as \emph{initDayLimit} and\emph{ initMulitowned}, can be called by anyone without authorization, the attacker is also able to control the multi-signature wallet by calling the \emph{initWallet} in \emph{WalletLibrary} to claim the ownership of the multi-signature wallet. Therefore, if the attacker destroys the wallet through the \emph{self-destruct} operation, this will freeze all the user's wallets that depend on this public multi-signature wallet.

\emph{(4) Beauty Chain Integer Overflow.}
The improper assignment operations of integer variables in smart contracts are prone to integer overflow. In the real world, there are many cases of integer overflow/underflow in smart contracts. For example, in the Proof-of-Week-Hands (POWH)~\cite{POWH}, around 2,000 Ether was stolen due to its integer overflow vulnerability. In recent years, the most famous smart contract integer overflow incident is the Beauty Chain vulnerability, in which attackers exploit the batch transfer method \emph{batchTransfer} in the contract to generate an unlimited number of BEC token, leading to the value of the BEC token evaporated to zero.

\begin{figure}
	\centering
	\includegraphics[width=8.6cm]{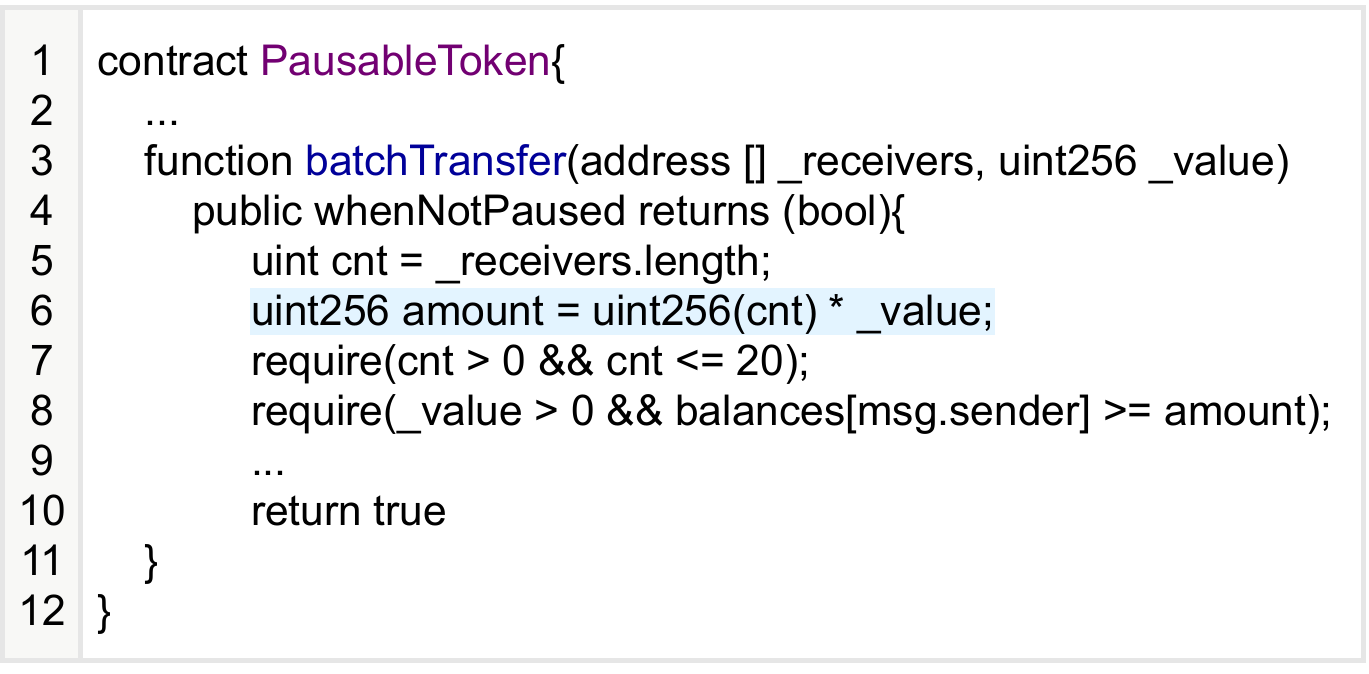}
	\caption{An example of Beauty Chain Integer Overflow Attack}
	\label{fig_BEC}
\end{figure}

Fig.~\ref{fig_BEC} presents the code snippets of \emph{batchTransfer} function in the BEC contract. \emph{batchTransfer} is a batch transfer function. Unfortunately, this function has an integer overflow vulnerability, which lies in line 6, \emph{i.e.,} ``{\small$uint256  amount = uint256 (cnt) * \_value$}'', where \emph{uint256} represents a 256-bit unsigned integer with a data range in [0,  $2^{256}$-1]. The attacker just exploits this nature by assigning a large value into \emph{\_value} that makes the \emph{amount} variable exceed the data range of \emph{uint256}, incurring an integer overflow vulnerability. Then, attackers copy an unlimited number of BEC tokens, thereby causing the value of the BEC token to evaporate to zero.

\begin{figure}
	\centering
	\includegraphics[width=8.6cm]{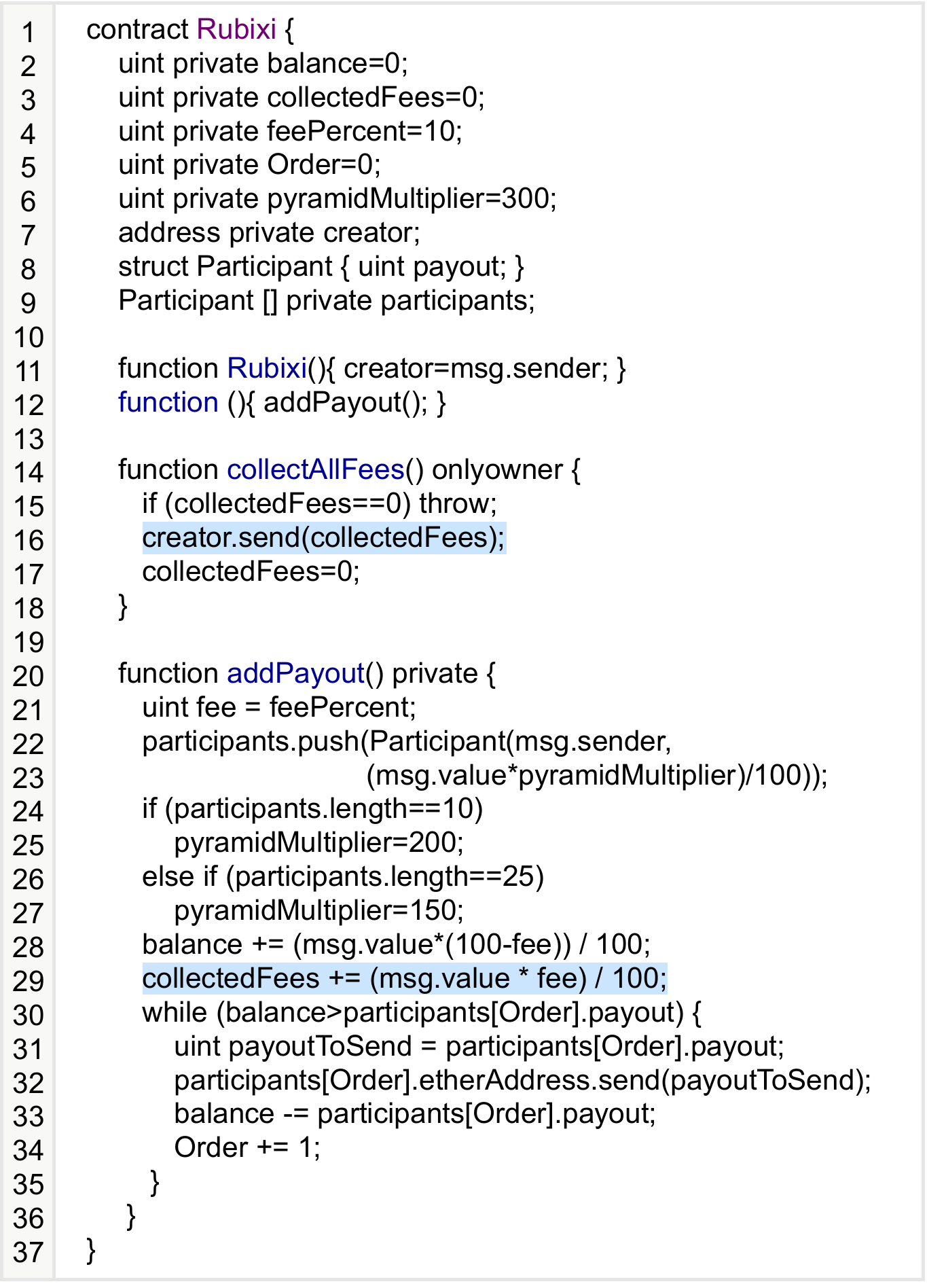}
	\caption{An example of Ponzi Scheme Attack -- Rubixi}
	\label{fig_rubixi}
\end{figure}

\emph{(5) Ponzi Scheme Rubixi.} 
A Ponzi scheme is a classic investment fraud method, which is typically characterized by paying existing investors a portion of the participation fee for new investors in return. The originator of a Ponzi scheme usually promises that joining an investment will generate high returns. The rewards are low and the risk is low to attract new investors. However, these types of scams ultimately only harm the interests of the vast majority of participants. 

With the widespread use of smart contracts, Ponzi schemes have gradually taken on new characteristics. Due to the anonymity of the blockchain, all participants cannot know the real identity of the contract initiator, resulting in many Ponzi schemes that can be disguised in smart contracts. under disguise. We call this type of Ponzi scheme a smart contract Ponzi scheme. Since smart contracts are self-executing and cannot be tampered with, it is one of the most beneficial means for Ponzi schemes to attract victims. In order to better understand the smart contract Ponzi scheme, this subsection presents a typical \emph{Rubixi} case in Fig.~\ref{fig_rubixi}.

The \emph{Rubixi} contract contains a constructor \emph{Rubixi}, a fallback function, a function \emph{collectAllFees}, and a function \emph{addPayout}, respectively. 
\begin{itemize}
\item The \emph{Rubixi} constructor is executed when the contract is created, and only once.
\item The fallback function (i.e. \emph{function () \{ addPayout(); \}}) will be automatically executed when an ether transfer is received, and the fallback function will be triggered automatically when the participant puts Ether into the \emph{Rubixi} contract.
\item The function \emph{collectAllFees} is used by the contract creator to withdraw funds deposited in the contract.
\item The \emph{addPayout} function is the most critical function, which implements the main logic of the Ponzi scheme: (1) Record the participant's address and participation fee; (2) Calculate the participant's investment fee; (3) When the balance in the contract is sufficient, then the remuneration fee is paid to the existing participants.
\end{itemize}

Obviously, it can be seen from the code snippet in Fig.~\ref{fig_rubixi}, \emph{pyramidMultiplier} is a key variable that controls how much profit a participant can earn. To attract early investment participants, the contract owner pre-sets its value to 300, when the number of participants reaches 10 and 25, the value of \emph{pyramidMultiplier} is reduced to 200 and 150, respectively. From the perspective of the entire contract logic, the main purpose of the contract initiator is to steal investment participants' investment fees, such as \emph{collectedFees+=(msg.value*fee)/100} (Line 29) is used to charge a 10\% participation fee for each investment, and the fee is extracted by calling the \emph{collectAllFees} function \emph{creator.send(collectedFees)} (Line 16).

\section{Vulnerability Detection Techniques for Smart Contracts}
\label{sec:analysis_method}
The security issues of smart contracts are becoming a significant concern for both researchers and developers. To prevent malicious attackers from exploiting smart contract vulnerabilities, researchers have tried various methods to comprehensively analyze the source code or bytecode of Ethereum smart contracts.

Traditional program vulnerability detection employs feature matching techniques~\cite{songming2020dc}, which extract the malicious code and analyzes the source code through the matching module. Unfortunately, this method restricts the application scope and incurs high false negatives. In recent years, there are five majority methods for smart contract vulnerability detection, including formal verification~\cite{bhargavan2016formal,bai2018formal,abdellatif2018formal}, symbolic execution \cite{mossberg2019manticore,shishkin2019debugging,zhao2020smart}, fuzzing detection~\cite{takanen2018fuzzing,liang2018fuzzing,wang2020systematic}, intermediate representation~\cite{zhao2012formalizing} and deep learning~\cite{wang2016automatically,shi2018vulnerability,wu2017vulnerability,li2018vuldeepecker,russell2018automated}.

\subsection{Formal Verification}
Formal verification is one of the significant technologies in security verification, which can transform the concepts, judgments, and ratiocination into a smart contract model through formal language, thereby eliminating the ambiguity and non-universality in the contract program. Moreover, this method verifies the correctness and safety of functions with rigorous logic and proof. Common formal verification methods include model checking and deductive verification. Model-checking lists all possible states and checks them individually to confirm whether the contract has the corresponding characteristic through state-space search. Deductive verification is based on the ideology of theorem-proof that uses logical formulas to describe the properties and proves the characteristics of the system through the evidence rules. Generally, formal verification takes advantage of mathematical logic to verify the code implementation whether satisfies some pivotal characteristics.

Currently, formal verification technology has been successfully applied to many fields with high-security requirements, such as nuclear power~\cite{deutch2003future} and aerospace~\cite{mouritz2012introduction}. By using formal methods for security analysis of smart contracts, we can ensure that contract generation is more standardized and contract execution is more reliable. To summarize, we conclude five formal verification methods as follows.

\begin{figure}
	\centering
	\includegraphics[width=8.65cm]{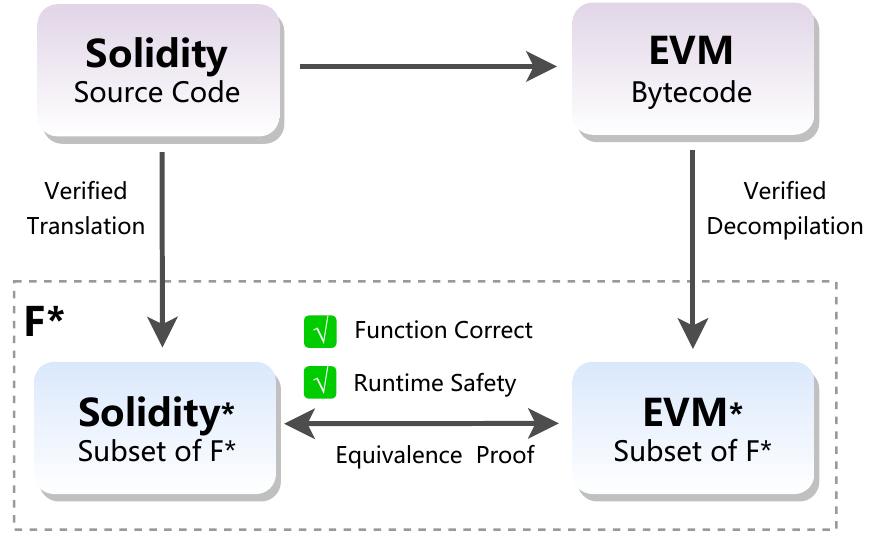}
	\caption{The overall architecture of \emph{F}* Framework}
	\label{fig_F}
\end{figure}

\emph{(1) F* Framework.}
F* framework is one of the formal frameworks developed by~\cite{grishchenko2018semantic}, which formalizes the semantics of the EVM bytecode and compiles the bytecode into \emph{Ocaml} form. Then, the smart contract source code and bytecode are converted into the functional programming language \emph{F}* for analyzing the security and verifying the correctness of functions during the contract runtime. Fig.~\ref{fig_F} depicts the overall architecture of the \emph{F}* framework, which implements the two modules, i.e., \emph{Solidity}* and \emph{EVM}*, to verify the functional equivalence between source code and bytecode, ensuring the correctness of the outputs.

\emph{(2) KEVM Framework.}
KEVM~\cite{hildenbrandt2018kevm} is a formal analysis framework, which utilizes the \textit{$\mathbb{K}$} framework to construct an executable formal specification based on the EVM bytecode stack. Further, KEVM serves as a platform for building a wide range of analysis tools and other semantic extensions for EVM.

\emph{(3) Isabelle/HOL.}
Isabelle/HOL~\cite{amani2018towards} is a proof assistant designed to infer and validate the correctness of EVM bytecode based on separation logic. This tool constructs the bytecode sequence into linear code blocks and splits the contract into basic blocks. Then, it further builds a logic program on this basis for reasoning verification.

\emph{(4) ZEUS.}
ZEUS~\cite{kalra2018zeus} is a static analysis tool, which can verify the correctness of smart contracts and validate their fairness. This method employs abstract interpretation, symbolic model checking, and constraint statements to quickly verify the security of smart contracts. In total, it can detect six security vulnerabilities in smart contracts including reentrancy, integer overflow/underflow, transaction order dependence, etc.

\emph{(5) VaaS.}
VaaS~\cite{garfatta2021survey} is a “one-click” smart contract security detection platform based on the formal verification method. While this tool is able to automatically detect 10 major items and 32 small items of conventional security vulnerabilities in smart contracts, it can also accurately and efficiently locate the risk code location and provide modification suggestions.

\subsection{Symbolic Execution}
The primary idea of symbolic execution is to symbolize variables in the program code, which maintains a set of constraints for all execution paths by symbolizing program input. With symbolic execution, the constraint solver is used to solve the constraints and determine the reason for execution input. Finally, developers can use the constraint solver to get a new test input to detect whether the symbol value has a potential vulnerability.

The execution process of symbolic execution applied to smart contract vulnerability detection can be classified into the following four steps: 1) symbolize the variable values in the contract; 2) explain the instructions in the execution program one by one; 3) update the execution status and collect path constraints to explore all executable paths in the program; 4) discover the corresponding security issues. In this subsection, we present eight symbolic execution methods.

\begin{figure}
	\centering
	\includegraphics[width=8.5cm]{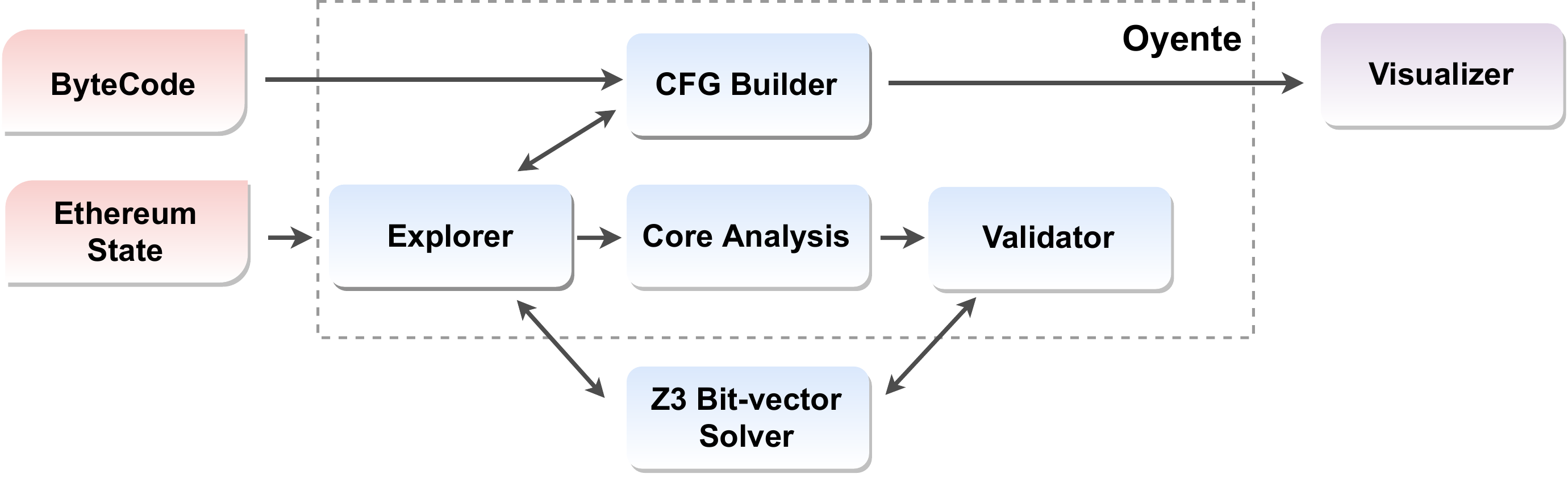}
	\caption{The overall architecture and execution process of Oyente}
	\label{fig_Oyente}
\end{figure}

\emph{(1) Oyente.}
Oyente~\cite{luu2016making} is one of the pioneer vulnerability detection tools for smart contracts, which utilizes symbolic execution to detect smart contract vulnerabilities based on the control flow graph (CFG), which takes the bytecode and the state of smart contracts as input to simulate EVM and traverse different execution paths of a certain contract. There are four modules in Oyente, including \emph{CFGBuilder}, \emph{Explorer}, \emph{CoreAnalysis}, and \emph{Validator}, and the overall architecture is illustrated in Fig.~\ref{fig_Oyente}.

\emph{(2) Mythril.}
Mythril~\cite{mueller2017framework} is a smart contract static analysis tool that combines concept analysis, taint analysis, and control flow verification to detect common vulnerabilities in Ethereum smart contracts, including reentrancy vulnerabilities, integer overflow, exception handling, etc.

\emph{(3) Osiris.}
Osiris~\cite{torres2018osiris} is a static analysis framework for smart contracts, which consists of three components, i.e., symbolic analysis, taint analysis, and integer error detection. This tool can detect three different types of integer errors, namely arithmetic errors, truncation errors, and signature errors. 

\emph{(4) Gasper.}
To monitor the gas consumption of smart contracts, Chen et al.~\cite{chen2017under}  present a static analysis tool named Gasper, which focuses on gas costly patterns from the existing smart contracts. Gasper takes the bytecode as the input to identify gas costly patterns, which runs symbolic execution on bytecode to find all the reachable code blocks in a smart contract. Specifically, this tool employs the \emph{Z3} solver~\cite{moura2008z3} to confirm whether the condition is true or false.

\begin{figure}
	\centering
	\includegraphics[width=8.4cm]{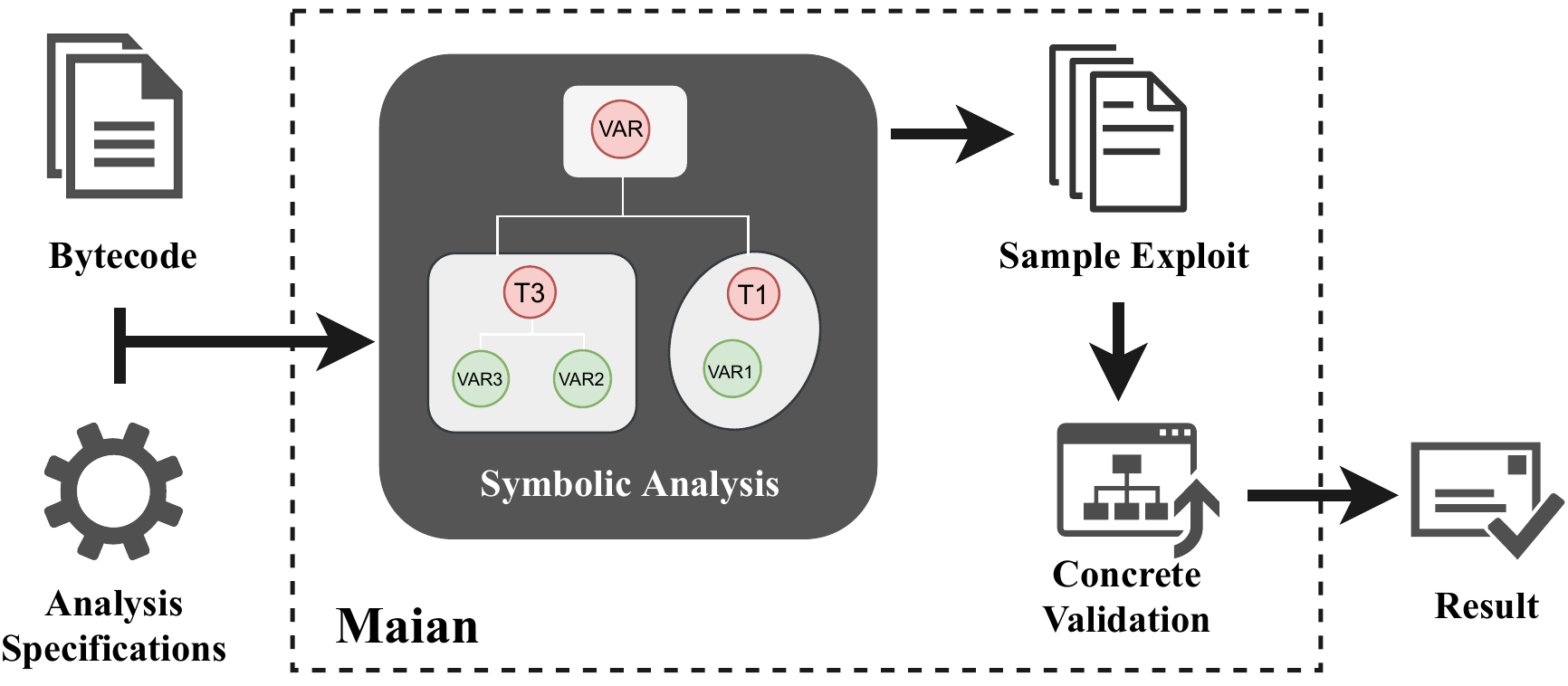}
	\caption{The overall architecture and execution process of Maian}
	\label{fig_Maian}
\end{figure}

\emph{(5) Maian.}
Maian~\cite{nikolic2018finding} detects smart contract vulnerabilities by using dynamic analysis, which discovers security vulnerabilities through a long sequence of invocations during the contract runtime. Fig.~\ref{fig_Maian} describes the overall architecture of Maian, which consists of two major components, namely symbolic analysis and concrete validation. Different from other detection tools, Maian focuses on the problematic smart contracts that can be labeled into three categories: \emph{Greedy}, \emph{Prodigal}, and \emph{Suicidal}. Specifically, asset frozen indefinitely is regarded as \emph{Greedy}, asset prone-leaked to unfamiliar accounts is looked as \emph{Prodigal}, and a contract destroyed arbitrarily by everyone is treated as \emph{Suicidal}.

\emph{(6) Securify.}
Securify~\cite{tsankov2018securify} is a static security analyzer for Ethereum smart contracts, which has the characteristics of scalability, fully automated, and high accuracy. Securify identifies the smart contacts vulnerabilities by analyzing the dependency graph and extracting precise semantic information from the bytecode, and checks compliance and violation patterns that capture sufficient conditions for proving if a property holds or not.

\emph{(7) TeEther.}
Different from traditional vulnerability detection tools, TeEther~\cite{krupp2018teether} considers the automatic identification of smart contract vulnerabilities and designs the method of contract generation. Based on the symbolic execution, this tool transverses the critical execution paths by analyzing the bytecode in order to solve the security issues in the contract.

\emph{(8) Sereum.}
Sereum~\cite{rodler2018sereum} is a novel detection solution that focuses on the reentrancy vulnerability of smart contracts. The tool employs dynamic taint tracking to monitor the data flow during the contract execution, thereby automatically avoiding inconsistent states and effectively preventing reentrant attacks.
\begin{figure}
	\centering
	\includegraphics[width=8.5cm]{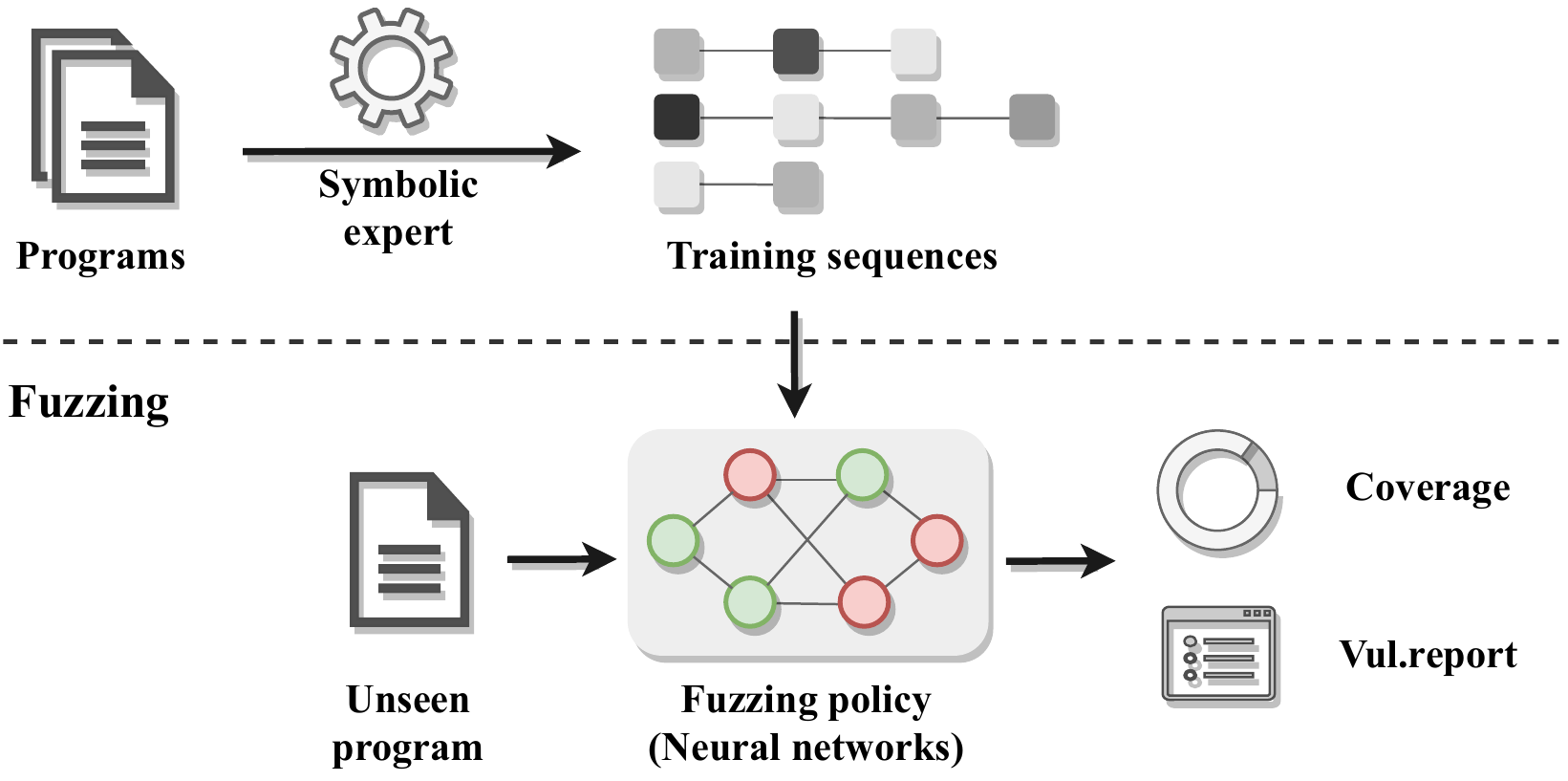}
	\caption{The overall architecture and execution process of ILF}
	\label{fig_ILF}
\end{figure}

\begin{figure*}
	\centering
	\includegraphics[width=16.9cm]{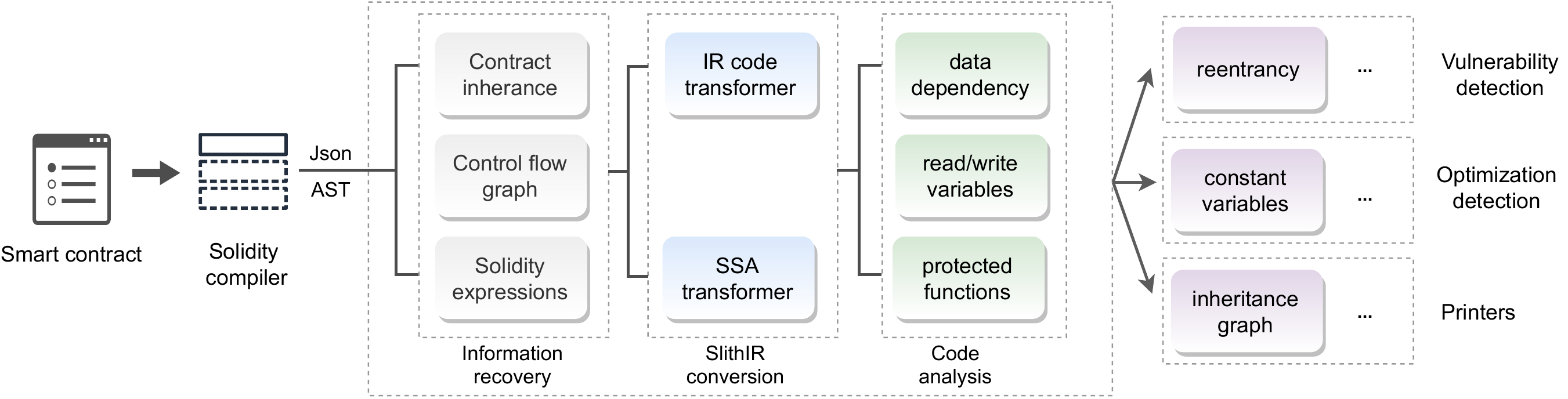}
	\caption{The overall architecture and execution process of Slither}
	\label{fig_Slither}
\end{figure*}

\subsection{Fuzzing Test}
Fuzzing test is one of the most popular software analysis and vulnerability detection techniques. Conceptually, the core idea of fuzzing test is to provide a large number of test samples for the program and monitor the abnormal behavior during the program execution. Compared with other testing technologies, fuzzing test is easy to be deployed and has good scalability and applicability. In this subsection, we introduce three fuzzing detection methods for smart contract vulnerability detection.

\emph{(1) ContractFuzzer.}
ContractFuzzer~\cite{jiang2018contractfuzzer} is the first fuzzing-based dynamic analysis method for detecting Ethereum smart contract security vulnerability, which generates fuzzing input based on the smart contract ABI specification and designs a test plan to detect vulnerabilities. First, ContractFuzzer configures the EVM and records the runtime behavior of the smart contract. Then, it detects vulnerabilities by analyzing these recorded logs. 

\emph{(2) Regurad.}
Regurad~\cite{liu2018reguard} is also a fuzzing analyzer, which focuses on the reentrancy vulnerabilities of smart contracts. Regurad performs the fuzzing test by iteratively generating random and diversifying test cases, thereby tracking the contract execution and further identifying the reentrancy vulnerability dynamically.

\emph{(3) ILF.}
ILF~\cite{he2019learning} is a neural network-based smart contract \emph{Fuzzer}, which is dedicated to generating better test cases and transaction sequences in the fuzzing of smart contracts. The solution of ILF is to first employ the symbolic execution engine to produce a large number of excellent call sequences, and then use the neural networks to learn the characteristics of these invocation sequences to guide the fuzzing engine to yield excellent scheduling strategies. 
Fig.~\ref{fig_ILF} illustrates the high-level idea of ILF, which trains a suitable architecture of neural networks that captures a probabilistic fuzzing policy for generating contract inputs. Then, ILF utilizes the learned policy to produce input sequences for fuzzing unseen contracts.

\subsection{Intermediate Representation}
To analyze smart contracts more accurately, researchers explore converting smart contract source code or bytecode into an intermediate representation (IR) with highly semantic information. They discover security issues by analyzing the intermediate representation of the contract. Here, we summarize the following six types of smart contract analysis tools using intermediate representation.

\emph{(1) Slither.}
Slither~\cite{feist2019slither} is a static analysis framework for Ethereum smart contract analysis to provide rich information, which combines data flow analysis and taint analysis. This tool converts the smart contract source code into an intermediate representation named \emph{SlithIR}, which employs a static single allocation (SSA) form and reduced instruction set to simplify the analysis process while retaining the lost semantic information when the source code is converted to EVM bytecode. Fig.~\ref{fig_Slither}  describes the core detection process of Slither, which is not only used to detect common vulnerabilities in smart contracts but also can give suggestions on contract code optimization.

\emph{(2) Vandal.}
Vandal~\cite{brent2018vandal} is a smart contract static analysis tool at the EVM bytecode level, consisting of an analysis pipeline and a decompiler. The decompiler performs abstract interpretation to convert the contract bytecode into a high-level intermediate representation (IR) in the form of logical relations, and then uses novel logic-driven methods to analyze security vulnerabilities.

\emph{(3) Madmax.}
Madmax~\cite{grech2018madmax} is a gas-oriented vulnerability analysis tool for Ethereum smart contracts, which performs control flow analysis and implements a decompiler program structure to detect smart contract security vulnerabilities based on Vandal. The tool similarly decompiles the EVM bytecode into the high-semantic intermediate representation but focuses on detecting gas-oriented vulnerabilities, \emph{e.g.,} Ether frozen.

\emph{(4) Ethir.}
Albert et al.~\cite{albert2018ethir} present a static analysis tool \emph{Ethir}, to analyze Ethereum smart contracts at the bytecode level on the basis of the control flow graph (CFG) generated by Oyente. Then, the tool converts the CFG into a rule-based intermediate representation (RBP) to analyze and infer the security properties of the EVM bytecode.

\begin{figure*}
	\centering
	\includegraphics[width=16cm]{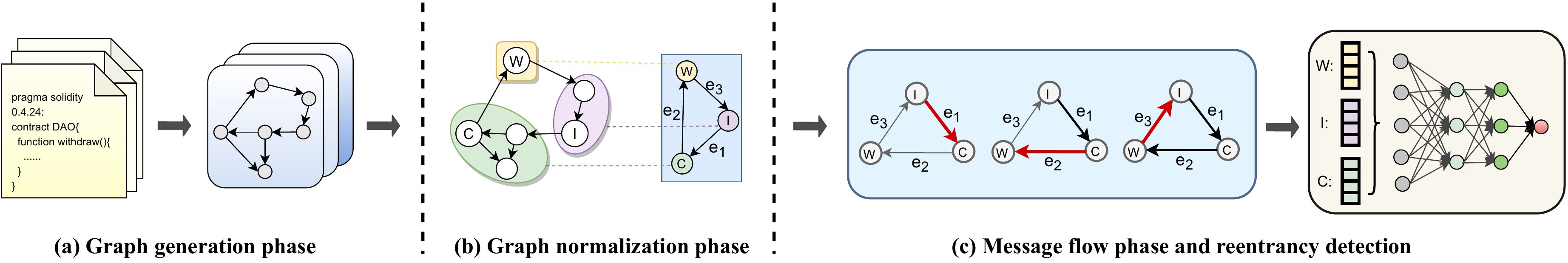}
	\caption{The overall architecture of TMP. (a) The contract graph generation phase, which constructs the graph from source code by extracting the control- and data- flow semantics in the code; (b) the graph normalization phase, which passes the features of normal nodes to their neighboring core nodes; (c) the novel temporal message flow network for vulnerability modeling and detection.}
	\label{fig_TMP}
\end{figure*} 

\emph{(5) Smartcheck.}
SmartCheck~\cite{tikhomirov2018smartcheck} is an extensible static analysis tool for smart contracts, which converts the source code into an XML-based intermediate representation. Then, SmartCheck detects the smart contract vulnerabilities in the middle representation of the contract based on the analysis of the XPath patterns. Moreover, this tool can be improved in multiple directions, such as improving the grammar, making patterns more precise, and adding new patterns.

\emph{(6) ContractGuard.}
ContractGuard~\cite{wang2019contractguard} is the first intrusion detection tool for Ethereum smart contracts against attacks. Based on the intrusion detection system (IDS), ContractGuard detects the abnormal control flow caused by potential attacks, which realizes intrusion detection by embedding the corresponding decentralized nature in the contracts to profile context-tagged acyclic path.

\subsection{Deep Learning}
In recent years, there has been an increasing practice of deep learning in the field of program security analysis~\cite{wang2016automatically,shi2018vulnerability,wu2017vulnerability,li2018vuldeepecker,russell2018automated}, which has achieved encouraging results. The advancement of deep learning technology promotes the birth of various security detection methods using deep learning models. Besides, deep learning-based methods have good expansibility and adaptability to new vulnerabilities. In this subsection, we review five recent kinds of research using deep learning models for smart contract vulnerability detection.

\emph{(1) SaferSC.}
SaferSC \cite{tann2018towards} is the first vulnerability detection model based on deep learning for smart contracts. This model focuses on three smart contract vulnerabilities defined by Maian and achieves a higher detection accuracy than Maian \cite{nikolic2018finding}. Besides, SaferSC analyzes the operation code (opcode) of the smart contract and employs the LSTM network to build a sequence model at the opcode level to achieve precise vulnerability detection. 

\emph{(2) RecChecker.}
RecChecker~\cite{qian2020towards} is a deep learning-based method, which focuses on detecting smart contract reentrancy vulnerability. This method captures the basic semantic information and control flow dependency information in a smart contract by converting a contract source code into the form of a \emph{contract snippet}. Then, RecChecker constructs the Bidirectional Long Short-Term Memory (BLSTM) with attention mechanism (Attention)~\cite{gilmer2017neural} to achieve the automatic detection of reentrancy vulnerability.

\emph{(3) DR-GCN.}
DR-GCN \cite{zhuang2020smart} is the first to explore the \emph{contract graph} for smart contract vulnerability detection, which uses the graph convolutional neural network (GCN) \cite{zhou2020graph} to construct a vulnerability detection model by converting a smart contract into a structure of \emph{contract graph} with a high degree of semantic representation. DR-GCN analyzes smart contracts on two platforms, \emph{i.e.,} Ethereum and VNT Chain, and detects three types of vulnerabilities, \emph{i.e.,}  reentrancy, timestamp dependency, and infinite loop.

\emph{(4) TMP.}
Temporal message propagation (TMP) is an advanced version of using \emph{contract graph} for smart contract vulnerability detection \cite{zhuang2020smart}. By constructing a \emph{contract graph} of a smart contract, the key functions and variables are converted into core nodes with high semantic information, while the execution modes are reflected into the directed edge depending on the control- and data- dependency. Furthermore, TMP considers the temporal information on the edges of the \emph{contract graph} and builds the temporal message propagation graph neural network \cite{gilmer2017neural,li2015gated}. As shown in Fig.~\ref{fig_TMP}, TMP consists of three phases: (1) a graph generation phase, which extracts the control flow and data flow semantics from the source code and explicitly models the fallback mechanism; (2) a graph normalization phase inspired by the $k$-partite graph, and (3) a novel temporal message flow network for vulnerability modeling and detection. 

\begin{table*}
	\centering
	\caption{Overview of different methods for smart contract vulnerability detection}
	\resizebox{0.98\textwidth}{!}{
		\begin{tabular}[htbp]{m{3cm}<{\centering} m{2cm}<{\centering} m{2.5cm}<{\centering} m{2.5cm}<{\centering}  m{6cm}<{\centering}  m{1cm}<{\centering} } 
			\toprule
			\textbf{Analysis Method}&\textbf{Detection Tool}&\textbf{Automatic Degree}&\textbf{Language Support}&\textbf{Public Available}&\textbf{Ref.}\\
			\hline
			\multirow{5}{*}{Formal verification}&F* framework&Semi-automatic&EVM, Ocaml&Not Available&\cite{grishchenko2018semantic}\\  
			&KEVM&Semi-automatic&Solidity, EVM&https://github.com/kframework/evm-semantics&\cite{hildenbrandt2018kevm}\\
			&Isabelle/HOL&Semi-automatic&EVM, Ocaml&https://github.com/pirapira/eth-isabelle&\cite{amani2018towards}\\
			&ZEUS&Fully automatic&Solidity, EVM&Not Available&\cite{kalra2018zeus}\\
			&VaaS&Fully automatic&Solidity, EVM&Not Available&\cite{garfatta2021survey}\\
			\hline
			\multirow{6}{*}{Symbolic execution}&Oyente&Fully automatic&Solidity, EVM&https://github.com/melonproject/oyente&\cite{luu2016making}\\ 
			&Mythril&Fully automatic&Solidity, EVM&https://github.com/ConsenSys/mythril&\cite{mueller2017framework}\\
			&Osiris&Fully automatic&Solidity, EVM&https://github.com/christoftorres/Osiris&\cite{torres2018osiris}\\
		    &Gasper&Fully automatic&Solidity, EVM&Not Available&\cite{chen2017under}\\
			&Maian&Fully automatic&Solidity, EVM&https://github.com/MAIAN-tool/MAIAN&\cite{nikolic2018finding}\\
			&Securify&Fully automatic&Solidity, EVM&https://github.com/eth-sri/securify2&\cite{tsankov2018securify}\\
			&TeEther&Fully automatic&Solidity, EVM&https://github.com/nescio007/teether&\cite{krupp2018teether}\\
			&Sereum&Fully automatic&Solidity, EVM&https://github.com/uni-due-syssec/eth-reentrancy-attack-patterns&\cite{rodler2018sereum}\\
			\hline
			\multirow{3}{*}{Fuzzing detection}&ContractFuzzer&Fully automatic&Solidity, EVM&https://github.com/gongbell/ContractFuzzer&\cite{jiang2018contractfuzzer}\\   
			&Regurad&Fully automatic&Solidity, EVM&Not Available&\cite{liu2018reguard}\\
			&ILF&Fully automatic&Solidity, EVM&https://github.com/eth-sri/ilf&\cite{he2019learning}\\
			\hline
			\multirow{6}{*}{Intermediate representation}&Slither&Fully automatic&Solidity, EVM&https://github.com/crytic/slither&\cite{feist2019slither}\\  
			&Vandal&Fully automatic&Solidity, EVM&https://github.com/usyd-blockchain/vandal&\cite{brent2018vandal}\\
			&Madmax&Fully automatic&Solidity, EVM&https://github.com/nevillegrech/MadMax&\cite{grech2018madmax}\\
			&Ethir&Fully automatic&Solidity, EVM&https://github.com/costa-group/ethIR&\cite{albert2018ethir}\\
			&Smartcheck&Fully automatic&Solidity, XML&https://https://github.com/smartdec/smartcheck&\cite{tikhomirov2018smartcheck}\\
			&ContractGuard&Fully automatic&Solidity&https://https://github.com/contractguard/experiments&\cite{wang2019contractguard}\\
			\hline
			\multirow{5}{*}{Deep learning}&SaferSC&Fully automatic&Solidity, EVM&https://github.com/wesleyjtann/Safe-SmartContracts&\cite{tann2018towards}\\  
			&RecChecker&Fully automatic&Solidity&https://github.com/Messi-Q/ReChecker&\cite{qian2020towards}\\
			&DR-GCN&Fully automatic&Solidity&https://github.com/Messi-Q/GraphDeeSmartContract&\cite{zhuang2020smart}\\
			&TMP&Fully automatic&Solidity&https://github.com/Messi-Q/GNNSCVulDetector&\cite{zhuang2020smart}\\
			&ContractWard&Fully automatic&Solidity&Not Available&\cite{wang2020contractward}\\
			&CGE&Fully automatic&Solidity&https://github.com/Messi-Q/GPSCVulDetector&\cite{Liu_2021}\\
			&AME&Fully automatic&Solidity&https://github.com/Messi-Q/AMEVulDetector&\cite{liu2021smart}\\
			\bottomrule
	\end{tabular}}
\label{table_detection tools}
\end{table*}

\emph{(5) ContractWard.}
ContractWard \cite{wang2020contractward} utilizes machine learning techniques to detect vulnerabilities in smart contracts, which learns the patterns of vulnerable contracts in training samples. ContractWard extracts bigram features from smart contract opcodes and employs a variety of machine learning algorithms and sampling algorithms, which can effectively and efficiently detect six types of vulnerabilities based on extracted static characteristics.

\emph{(6) CGE.}
CGE\cite{Liu_2021}  is the first to investigate the idea of fusing conventional expert patterns and graph-neural-network extracted features for smart contract vulnerability detection. CGE proposes to characterize the contract function source code as contract graphs. Then, CGE explicitly normalizes the graph to highlight key variables and invocations. Finally, CGE utilizes a novel temporal message propagation network to automatically capture semantic graph features and combines the graph feature with designed expert patterns to yield a final detection.

\emph{(7) AME.}
AME\cite{liu2021smart} is a new system beyond pure neural networks that can automatically detect vulnerabilities and incorporate expert patterns into networks in an explainable fashion. Firstly, AME extracts expert patterns of a specific vulnerability from the function code. Second, AME utilizes a graph construction and normalization module which transforms the function code into the code semantic graph vector. Finally, AME combines local expert patterns and the global graph feature for vulnerability detection and outputs explainable weights. AME is an important tool for explainable and accurate contract vulnerability detection.

\renewcommand{\arraystretch}{1.2}
\section{Evaluation}
\label{sec:tools}
In this section, we first overview 29 different smart contract detection methods. Then, we analyze these methods from different aspects and compare how well they support detecting various types of vulnerabilities. Finally, we evaluate the performance of these tools in terms of accuracy, F1-Score, and average detection time.

\subsection{Overview of Existing Methods}
Table~\ref{table_detection tools} displays different smart contract detection methods, in which the first column represents five effective smart contract analysis methods, while the second column lists the corresponding detection tools. In the third column, we describe the automatic degree of each detection tool. Here, \emph{fully automatic} refers to the end-to-end solution, which means that a tool takes a contract as input and outputs the specific vulnerability detection results, and \emph{semi-automatic} represents manually defining relevant contract attributes during the detection process. For example, formal verification methods employ theorems to prove the security of smart contracts. Since these proofs are semi-automated, formal verification methods require a lot of manual operations to conduct smart contract verification and analysis. The fourth column summarizes the smart contract languages and forms supported by the corresponding detection tools, and the fifth column presents the degree of open source with the open-source addresses. In the last column, we list the references for each method.

According to the statistical results in Table~\ref{table_detection tools}, we present the overall analysis as follows.
\begin{itemize}
	\item Compared with other methods, formal verification has a low degree of automated and open-sourced nature.
	\item There are many kinds of detection tools based on symbolic execution and intermediate representation, and all of them can perform fully automatic vulnerability detection. Most of them have open-source code.
	\item There are few detection tools based on the fuzzing test. The reason may be that the implementation and operation of dynamic fuzzing detection methods are more complicated and cumbersome. Furthermore, due to the randomness of test cases, the methods of fuzzing detection cannot cover all the test paths.
	\item Most of the smart contract vulnerability detection methods based on deep learning still focus on the level of the contract source code. They can not only perform fully automatic detection but also with a high degree of open-sourced nature.
\end{itemize}

\renewcommand{\arraystretch}{1.35}
\begin{table*}
	\caption{The overview of detectable vulnerability supported by existing vulnerability detection methods}
	\resizebox{1\textwidth}{!}{
		\centering
		\begin{tabular}{|m{4cm}<{\centering}|
				m{0.8cm}<{\centering}|m{0.8cm}<{\centering}|m{0.8cm}<{\centering}|m{0.8cm}<{\centering}|m{0.8cm}<{\centering}|m{0.8cm}<{\centering}|m{0.8cm}<{\centering}|m{0.8cm}<{\centering}|m{0.8cm}<{\centering}|m{0.8cm}<{\centering}|m{0.8cm}<{\centering}|m{0.8cm}<{\centering}|m{0.8cm}<{\centering}|m{0.8cm}<{\centering}|m{0.8cm}<{\centering}|m{0.8cm}<{\centering}|}
			\hline
			\multirow{2}{*}{\textbf{Detection Tool}}
			&\multicolumn{9}{c|}{\textbf{Solidity code layer}}
			&\multicolumn{4}{c|}{\textbf{EVM execution layer}}		&\multicolumn{3}{c|}{\textbf{Block dependency layer}}\\
			\cline{2-17}
			&\rotatebox{90}{\textbf{Reentrancy}}
			&\rotatebox{90}{\textbf{Integer Overflow}}
			&\rotatebox{90}{\textbf{Access Control}}
			&\rotatebox{90}{\textbf{Exception Handling}}
			&\rotatebox{90}{\textbf{Denial of Service}}
			&\rotatebox{90}{\textbf{Type Mismatch}}
			&\rotatebox{90}{\textbf{Unknown Function Call}}
			&\rotatebox{90}{\textbf{Ether Frozen}}
			&\rotatebox{90}{\textbf{Replay Attack}}
			&\rotatebox{90}{\textbf{Short Address}}
			&\rotatebox{90}{\textbf{Ether Loss}}
			&\rotatebox{90}{\textbf{Call-Stack Overflow}}
			&\rotatebox{90}{\textbf{Tx.origin}}
			&\rotatebox{90}{\textbf{Timestamp Dependency}}
			&\rotatebox{90}{\textbf{Block Parameter Dependency}}
			&\rotatebox{90}{\textbf{Transaction Ordering Dependency}}\\
			\hline
			F* framework&\Checkmark&&&\Checkmark&&&\Checkmark&&&&&& &\Checkmark&\Checkmark&\Checkmark\\
			\hline
			ZEUS&\Checkmark&\Checkmark&&&&&&&\Checkmark&&&&&\Checkmark&\Checkmark&\Checkmark\\
			\hline
			VaaS&\Checkmark&\Checkmark&&\Checkmark&\Checkmark&&\Checkmark&&&&&&\Checkmark&\Checkmark&\Checkmark&\\
			\hline
			Oyente&\Checkmark&\Checkmark&&\Checkmark&&&&&\Checkmark&&&\Checkmark&&\Checkmark&&\Checkmark\\
			\hline
		    Mythril&\Checkmark&\Checkmark&&\Checkmark&\Checkmark&&&&\Checkmark&&\Checkmark&&\Checkmark&\Checkmark&\Checkmark&\\
	        \hline		
			Osiris&\Checkmark&\Checkmark&&\Checkmark&&&&&\Checkmark&&&\Checkmark&&\Checkmark&&\Checkmark\\
			\hline
			Maian&\Checkmark&&&&&&&\Checkmark&&&\Checkmark&\Checkmark&&&&\\
			\hline
			Securify&\Checkmark&&&\Checkmark&&\Checkmark&\Checkmark&\Checkmark&\Checkmark&\Checkmark&\Checkmark&\Checkmark&\Checkmark&\Checkmark&\Checkmark&\Checkmark\\
			\hline
			TeEther&\Checkmark&&&&&&\Checkmark&\Checkmark&&&&&&&&\\
			\hline
			Sereum&\Checkmark&&&&&&&&&&&&&&&\\
			\hline
			ContractFuzzer&\Checkmark&&&\Checkmark&&&\Checkmark&\Checkmark&&&&&&\Checkmark&\Checkmark&\\
			\hline
			Reguard&\Checkmark&&&&&&&&&&&&&&&\\
			\hline
			ILF&\Checkmark&&&\Checkmark&&&&\Checkmark&&&\Checkmark&&&\Checkmark&\Checkmark&\\
			\hline
			Slither&\Checkmark&&&&&\Checkmark&\Checkmark&\Checkmark&&&&\Checkmark&\Checkmark&\Checkmark&&\\
			\hline
			Vandal&\Checkmark&&&\Checkmark&&&&\Checkmark&&&\Checkmark&&\Checkmark&&&\Checkmark\\
			\hline
			Madmax&&\Checkmark&&&&&&\Checkmark&&&&&&&&\\
			\hline
			Ethir&\Checkmark&&&\Checkmark&&&&&&&&&&\Checkmark&&\Checkmark\\
			\hline
			Smartcheck&\Checkmark&\Checkmark&\Checkmark&&\Checkmark&&\Checkmark&\Checkmark&&&& &\Checkmark&\Checkmark&&\\
			\hline
			ContractGuard&\Checkmark&\Checkmark&&&\Checkmark&\Checkmark&\Checkmark&&&\Checkmark&&&\Checkmark&\Checkmark&\Checkmark&\Checkmark\\
			\hline
			SaferSC&\Checkmark&&&&&&&\Checkmark&&&\Checkmark &\Checkmark&&&&\\
			\hline
			RecChecker&\Checkmark&&&&&&&&&&&&&&&\\
			\hline
			DR-GCN&\Checkmark&&&&&&&&&&&&&\Checkmark&&\\
			\hline
			TMP&\Checkmark&&&&&&&&&&&&&\Checkmark&&\\
			\hline
			ContractWard&\Checkmark&\Checkmark&&&&&&&&&&\Checkmark&&\Checkmark&&\Checkmark\\
			\hline
			CGE&\Checkmark&&&&&&&&&&&&&\Checkmark&&\\
			\hline
			AME&\Checkmark&&&&&&&&&&&&&\Checkmark&&\\
			\hline
			KEVM framework&\multicolumn{16}{c|}{\multirow{2}*{KEVM framework and Isabelle/HOL provide verification conditions for contract program analysis and formal verification methods. The functional}}\\
			\cline{1-1}
			Isabelle/HOL&\multicolumn{16}{c|}{correctness and program logical rationality are verified during contract execution, but it is not used to detect specific contract vulnerabilities.}\\
			\hline
			Gasper&\multicolumn{16}{c|}{Gasper is a tool for automatically locating gas costly patterns by analyzing smart contracts’ bytecode.}\\
			\hline
		\end{tabular}
	}
	\label{table_vulnerability categories}
\end{table*}

\subsection{Result Analysis}
Table~\ref{table_vulnerability categories} summarizes 29 smart contract vulnerability detection tools and the detectable vulnerabilities supported by them, which are concluded into three categories. We elaborate on the specific analysis of the detectable vulnerabilities and performance of existing methods in detail.

\subsubsection{Comparison of existing methods from different levels} 
From the analysis of Solidity code layer, most of the detection tools support verifying the reentrancy vulnerability. Since the most famous \emph{The DAO incident} is caused by the reentrancy vulnerability, most researchers and developers first focus on the analysis of such vulnerability. In addition, many tools pay close attention to the integer overflow, exception handling, unknown function call, and Ether frozen, while these security vulnerabilities have incurred major security incidents, such as Beauty Chain integer overflow and Parity multi-signature wallet frozen. Further, we need to point out that there are few tools for analyzing access control, denial of service, type mismatch, and replay attack, while these four vulnerabilities occur less frequently and are easy to prevent.

From the analysis of the EVM execution layer, there are few tools for detecting the short address vulnerability, which can be attributed to the low probability of occurrence and ease of verification. Besides, both methods based on formal verification and deep learning fail to detect smart contract vulnerabilities at the EVM execution level, such as ZEUS, \emph{F}* framework, RecChecker, DR-GCN, and TMP.

From the analysis of Block dependency layer, timestamp dependency is a common smart contract vulnerability and easy to detect. Most detection tools support detecting timestamp dependency, and it is worth noting that \emph{F}* framework, ZEUS, ContractGuard, and Securify are able to detect all the vulnerabilities in the block dependency layer.

In summary, although most vulnerability types can be detected by corresponding tools, some easy-to-verify vulnerabilities are only supported by a few detection tools. For example, only two tools can detect access control and short address vulnerabilities, while the probability of these two vulnerabilities is relatively low, but the losses caused by such vulnerabilities are also inestimable. Therefore, comprehensive coverage for various vulnerabilities is still one of the urgent problems to be solved by these automated detection tools. Currently, with the rapid growth of the number of smart contracts, the numbers and types of smart contract vulnerabilities are also increasing. Using automated detection tools to conduct more comprehensive and scalable detection for smart contracts is a key issue that deserves to be further studied.

\subsubsection{Comparison of detectable vulnerabilities of existing methods} 
According to the statistics in Table~\ref{table_vulnerability categories}, formal verification methods detect few contract vulnerabilities. Among them, KEVM and Isabelle/HOL can support contract program analysis and verify functional correctness and program logical rationality of smart contracts during execution, while they are unable to detect specific contract vulnerabilities. Moreover, \emph{F}* framework, ZEUS, and VaaS do not support detecting vulnerabilities in the EVM execution layer. It is worth to point that most formal verification methods employ mathematical theorem proofs and complex mechanisms for verification, which are not easy to use.

Most of the detection tools based on symbolic execution can detect more kinds of contract vulnerabilities. For example, Oyente, Mythril, and Securify can respectively detect 7, 9, and 13 vulnerabilities, of which Securify can support detecting the most types among the 27 detection tools. Osiris, Gasper, Maian, TeEther, and Sereum detect fewer vulnerabilities, while Maian is designed to solve three unique vulnerabilities (Greedy, Prodigal, Suicidal), and Sereum only focuses on the reentrancy vulnerability.

Compared with other methods, ContractFuzzer, Reguard, and ILF based on fuzzing detection can only detect a few vulnerabilities. ContractFuzzer and ILF both support detecting 6 vulnerabilities, while Reguard only focuses on the reentrancy vulnerability.

Moreover, the detection tools based on the intermediate representation have achieved good results. Among them, Vandal, Slither, Smartcheck, and ContractGuard support detecting 6, 7, 8, and 10 vulnerabilities respectively, while Madmax and Ethir can only detect 2 and 4 vulnerabilities.

It is worth mentioning that deep learning-based methods are also for few vulnerabilities detection. SaferSC and ContractWard can detect 4 and 5 vulnerabilities respectively, while RecChecker only focuses on the reentrancy vulnerabilities. DR-GCN and TMP are both constructed for detecting reentrancy and timestamp dependency vulnerabilities.

To summarize, vulnerabilities covered by various detection tools are still incomplete. Most of them can only detect low-level security violations and vulnerabilities, and lack inferences during contract execution, making it difficult to detect external security issues caused by contract invocations. Therefore, in the context of the current growing number of smart contracts, it is still challenging to use a single detection tool to fully verify smart contract vulnerabilities.

\subsubsection{Performance comparison of existing methods}
We compare and analyze the performance of different vulnerability detection tools in detail shown in Table~\ref{table_detection results}. First, we select the representative detection tools from the five smart contract security analysis methods described in section~\ref{sec:analysis_method}, namely VaaS, Oyente, Smartcheck, Contract Fuzzer, and TMP. Then, we randomly collect 300 Ethereum smart contracts as test samples from the official website Etherscan\cite{Etherscan}. The performance of detection tools is evaluated from three aspects, i.e., accuracy, F1-Score, and average detection time, and we focus on three smart contract vulnerabilities, reentrancy, integer overflow, and timestamp dependency.  

\renewcommand{\arraystretch}{1.1}
\begin{table*}
	\caption{Performance comparison in terms of \emph{Accuracy, F1-Score, and Average Detection Time}, ‘—’ denotes not applicable}
	\centering
	\resizebox{0.94\textwidth}{!}{
		\begin{tabular}[htbp]{m{2.6cm}<{\centering} m{2.5cm}<{\centering}m{2.4cm}<{\centering}m{2.4cm}<{\centering}  m{3.3cm}<{\centering}} 
		\toprule
		\textbf{Vulnerability Type}&\textbf{Detection Tool}&\textbf{Accuracy (\%)}&\textbf{F1-Score (\%)}&\textbf{Average Detection Time (s)}\\
		\hline
		\multirow{5}*{Reentrancy}&VaaS& 82.54&73.95&159.4\\  
		&Oyente &61.62&44.96&29.6\\
		&Smartcheck&52.97&30.10&14.5\\
		&ContractFuzzer&67.89&52.67&352.2\\
		&TMP&84.48&74.15&2.5\\
		\hline
		\multirow{5}*{Integer overflow}&VaaS&86.80&80.10&159.4\\  
		&Oyente&66.85&59.3&29.6\\
		&Smartcheck&58.48&54.96&14.5\\
		&ContractFuzzer&—&—&—\\
		&TMP&—&—&—\\
		\hline
		\multirow{5}*{Timestamp dependency}&VaaS&89.20&82.46&159.4\\  
		&Oyente&59.45&41.53&29.6\\
		&Smartcheck&51.32&40.18&14.5\\
		&ContractFuzzer&68.08&52.49&352.2\\
		&TMP&83.45&79.19&2.1\\
		\bottomrule
	\end{tabular}}
\label{table_detection results}
\end{table*}

\textbf{(1) Accuracy.} To evaluate the pros and cons of detection tools, we first focus on the most common evaluation indicator, namely accuracy. Generally, we determine a classifier whether is effective in a sense through accuracy, which can objectively reflect the most direct effect of various detection tools. Vulnerability detection is actually a binary classification problem, that is, the detection tool predicts whether there is a certain vulnerability in a contract. For the binary classification problem, the matching result is usually used as an important evaluation indicator, including the following four situations:
\begin{itemize}
	\item True positive (TP). For a contract, the detection result is true and the real value is also true, which implies the detection result is correct.
	\item False positive (FP). For a contract, the detection result is true and the real value is also false, which implies the detection result has a false positive.
	\item False negative (FN). For a contract, the detection result is false and the real value is also true, which implies the detection result has a false negative.
	\item True negative (TN). For a contract, the detection result is false and the real value is also false, which implies the detection result is correct.
\end{itemize}

In this experiment, the result of TP+FP+FN+TN is equal to the amount of smart contract test samples, 
and the calculation of accuracy is denoted in formula~\ref{eq:eq1}.

\begin{equation}
\label{eq:eq1}
Accuracy =  \frac {TP + TN}{TP + FP + FN + TN}
\end{equation}

According to the evaluation results in Table~\ref{table_detection results}, TMP achieves the highest accuracy (84.48\%) in the reentrancy vulnerability detection, while VaaS and ContractFuzzer have the accuracy of 82.54\% and 67.89\% respectively. In contrast, the accuracy of Oyente and Smartcheck is slightly insufficient, only 61.62\% and 52.97\% respectively. For integer overflow vulnerability, the accuracy of VaaS is as high as 86.80\%, while Oyente and Smartcheck only have 66.85\% and  58.48\%, ContractFuzzer and TMP do not support detecting such vulnerability. For timestamp dependency, VaaS and TMP achieve higher detection accuracy of 89.20\% and 83.45\% respectively, while the detection accuracy of Oyente, Smartcheck, and ContractFuzzer are very low, only 59.45\%, 51.32\%, and 68.08\% respectively.

\textbf{(2) F1-Score.} F1-Score is an important measurement indicator in the binary classification problem. It is the harmonic average of precision and recall, which is usually treated as the final evaluation standard for some classification missions. The calculation of F1-Score is shown in formula (\ref{eq:eq2}-\ref{eq:eq4}). 
\begin{equation}
\label{eq:eq2}
Precision = \frac {TP}{TP + FP}
\end{equation}
\begin{equation}
\label{eq:eq3}
Recall = \frac {TP}{TP + FN}
\end{equation}
\begin{equation}
\label{eq:eq4}
F_{1} = 2 * \frac{Precision * Recall}{Precision + Recall}
\end{equation}

According to the evaluation results of Table~\ref{table_detection results}, TMP achieves the highest F1-Score (74.15\%) in the detection of reentrancy vulnerability, followed by VaaS (73.95\%), and the remaining detection tools are relatively low. For integer overflow vulnerability, VaaS achieves an F1-Score of 80.10\%, while Oyente and Smartcheck get the F1-Score of 59.64\% and 54.96\% respectively. For timestamp dependency, VaaS and TMP both obtain good F1-Score of 82.46\% and 79.19\%.

\textbf{(3) Average detection time.} Average detection time is also one of the significant indicators for evaluating automated detection tools. Currently, the long audit time of most detection tools leads to the low efficiency of vulnerability analysis. According to the evaluation results in Table~\ref{table_detection results}, the average detection time of VaaS and ContractFuzzer is 159.4 seconds and 352.2 seconds, respectively. In comparison, the average detection time consumed by Oyente and Smartcheck is 29.6 seconds and 14.5 seconds respectively. It is worth noting that TMP has a floating detection time for various vulnerabilities due to its different models for different vulnerabilities. For example, TMP takes 2.5 seconds for detecting reentrancy vulnerability, while only needs 2.1 seconds for timestamp dependency. Based on the above analysis, we summarize these five detection tools as follows.
\begin{itemize}
	\item  \textbf{VaaS} achieves high accuracy and F1-Score in the detection of the three vulnerabilities, but its average detection time is relatively long. VaaS is a ``one-click'' formal verification platform, which employs a variety of formal verification methods and has the characteristics of high accuracy, verification efficiency, and automation.
	\item  \textbf{Oyente} is a contract analysis tool based on symbolic execution, which has an insufficient detection degree on the three vulnerabilities. Oyente has a high probability of leading to high false-negative and false-positive by simplifying the loop processing in the contract and the judgment based on rule matching. 
	\item \textbf{Smartcheck} is a static tool that uses the XML-based intermediate representation to express and analyze smart contract security issues. However, it relies on inherent and simple logic rules, which cause high false-positive for smart contract vulnerability detection, resulting in low accuracy and F1- Score. It is worth mentioning that it takes relatively little time to detect vulnerabilities due to depending on rigid rules.
	\item \textbf{ContractFuzzer} is a smart contract security vulnerability fuzzing tool based on the Ethereum platform. Experimentally, it only supports detecting reentrancy and timestamp dependency vulnerabilities. Since fuzzing use cases have a limit on covering all the execute paths, it cannot achieve the ideal path coverage. In addition, ContractFuzzer runs on the Ethereum platform so it also needs to get a response from Ethereum when conducting the detection, which takes a lot of time to perform one detection.
	\item \textbf{TMP} is a novel smart contract vulnerability detection model based on a graph neural network, which has the characteristics of high scalability, accuracy, and batch detection. Technically, TMP can detect reentrancy and timestamp dependency, achieving encouraging results. Due to using the pre-trained detection model, the average detection time of TMP is very low. Moreover, TMP is the first to explore combing deep learning technology and employing graph neural networks in smart contract vulnerability detection, which greatly improves efficiency and accuracy.
\end{itemize}

\subsection{Evaluation and Improvement} 
\subsubsection{Limitation Analysis}
Although current smart contract vulnerability detection methods can detect smart contract vulnerabilities, they still have inherent limitations. This subsection specifically analyzes and discusses the aforementioned five types of vulnerability detection methods.
\begin{itemize}
\item  The formal verification method uses some mathematical means to deduce and prove the smart contract in its life cycle, which requires interactive verification and judgment, so the degree of automation is low. Moreover, it relies on manual secondary verification, which makes it incompatible with EVM Execution layer vulnerability. At the same time, because the formal verification method relies on rigorous mathematical derivation and verification, it cannot perform dynamic analysis. Thus, it lacks the detection and judgment of the executable path in the contract, resulting in a high false positive rate and leakage. For example, F* framework and KEVM convert smart contract bytecode into formal models and verify various properties in the contract code to detect vulnerabilities, they are still semi-automatic. ZEUS and VaaS have well-implemented fully automatic formal verification, but the vulnerabilities detected by them do not have a reachable execution path, producing a high false positive rate.
\item The symbolic execution method uses symbols to replace specific execution program instructions, collect path constraints, and traverse all executable paths in the contract program. Although this method effectively improves the detection effect of symbolic execution, it also significantly increases the computational resources and time overhead in the vulnerability analysis process. In addition, they cannot completely solve the problems of state space explosion and exponential growth of execution paths. For example, Oyente and Maian limit the number of loop conditions to improve efficiency in order to prevent the problem of path explosion, but it also leads to high leakage. It is worth pointing out that many symbolic execution methods cannot be fully automated, and also require human assistance and feedback.
\item Fuzzing methods largely rely on well-designed test cases, which monitor the abnormal behavior of contracts during dynamic execution. However, fuzzing has limited insight into the specific semantic code that leads to vulnerabilities, which makes it difficult to It is difficult to track down the exact code location where the vulnerability exists. For example, ContractFuzzer effectively reduces the false positive rate, but cannot achieve the ideal path coverage due to the randomness of its test case generation, making it difficult to find all potential threats.
\item Intermediate representations use control flow, data flow, and taint analysis to review contracts by converting the original smart contract into a corresponding intermediate representation, but they often rely on predefined semantic rules or analysis lists, making it impossible to detect smart contracts that have complex business logic. In addition, they cannot traverse the execution paths that may exist in the contract. For example, Slither’s intermediate representation SlithIR relies on fixed semantic rules and lacks formal semantics, which is limited to performing more detailed security analysis, so it cannot accurately detect the corresponding vulnerabilities. Smartcheck relies on rigid and simple predefined rules, so it cannot detect some contract vulnerabilities verified by taint analysis or dynamic execution.
\item The deep learning method usually preprocesses smart contracts to construct a data set that is conducive to model learning. For example, the literature~\cite{qian2020towards} uses the LSTM model to process the source sequence fragments of the smart contracts, and the literature~\cite{zhuang2020smart} processes the smart contracts through the GNN model. However, on one hand, these methods cannot highlight the key variables in the source code of smart contracts, resulting in insufficient semantic modeling and unsatisfactory detection results. On the other hand, due to the black-box nature of neural networks, their interpretability is poor in most cases, that is, they cannot give the exact location or code line where there may be a vulnerability like traditional detection tools. For example, TMP is an end-to-end vulnerability detection model, with contract testing Set as input, output the corresponding vulnerability detection results, its intermediate processing flow is a black box, so its interpretability is poor, and the detection results are unconvincing.
\end{itemize}

\subsubsection{Research challenges and ideas for improvement}
In view of the problems of the existing smart contract vulnerability detection methods, this section discusses and analyzes the research challenges and improvement ideas they face, mainly focusing on the following five aspects.

(1) Most of the existing formal verification technology research work is not highly automated, and the detected vulnerabilities may not have an accessible program path. Current formal verification methods use mathematical derivation to analyze contracts that may have complex vulnerabilities, which can be difficult. In addition, the detection of broader smart contract vulnerabilities using formal verification techniques still faces serious challenges. Future research should customize the corresponding verification for different vulnerability detection target specification descriptions, break through the technical limitations such as its inability to adapt to large-scale contracts and multiple vulnerability types, and expand the application scope of formal verification. Realize the transition from verifying general functional attributes and security attributes, and detecting common vulnerabilities to gradually solving the analysis and verification of smart contract vulnerabilities in complex business logic in business scenarios.

(2) The main challenge of symbolic execution currently is the problem of state space explosion and the exponential growth of execution paths. A feasible method in the future is to combine the audit experience of existing symbolic execution tools and vulnerability analysis to find high-risk smart contracts that are prone to vulnerabilities. Instructions, such as \texttt{SUICIDE}, \texttt{CALL}, \texttt{DELEGATECALL}, \texttt{ORIGIN}, \texttt{ASSERT}, define the paths involving these opcodes as key paths. In order to improve the efficiency of vulnerability detection, it is not necessary to check all possible execution paths in a specific practice, only symbolically execute the focused paths and perform vulnerability verification, thereby reducing the path space.

(3) Compared with traditional applications, smart contracts have many unique variables and functions, which brings new challenges to fuzzing smart contracts.

First, due to the characteristics of the state updating of smart contracts, it is extremely difficult to generate effective test cases. The traditional program fuzzing scheme only considers a single test case when generating test cases, so it is not suitable for smart contracts. Second, smart contracts run on virtual machines, and the reasons for their vulnerabilities are quite different from traditional programs. They neither cause program crashes nor do they have many common features for vulnerability detection. The origin of these vulnerabilities may come from different levels such as blockchain, virtual machines, and high-level language, and there are many differences between them, which also brings great challenges to the vulnerability detection of smart contracts.

Specifically, fuzzing relies on the robustness of its test cases, so it is necessary to further improve the existing test case generation algorithms. For example, using multi-objective optimization algorithms. In addition, fuzzing can also consider combining other detection methods to improve detection efficiencies, such as adopting a strategy that combines static analysis or symbolic execution. For example, static analysis is used to extract critical paths, and test cases are generated through symbolic execution, thereby improving the efficiency of fuzzing.

(4) Intermediate representation method usually converts smart contract source code or bytecode into a unique intermediate representation and then detects certain types of vulnerabilities based on the IR. At the same time, they also rely on vulnerability rules defined by experts. But these rules are often rigid and simple, which are easy to be exploited by attackers. Therefore, in order to improve the expansibility and adaptability of this kind of vulnerability detection method, researchers should focus on making the intermediate representation of smart contracts more universal, so that the unified representation of different smart contracts can be considered while detecting various types of vulnerabilities. In addition, the combination of static analysis and dynamic execution is an effective method to improve the accuracy of vulnerability detection. Currently, most of the detection methods based on intermediate representations are static analysis, which lacks the use of dynamic execution for verification. Therefore, this is not only the key challenge currently faced by intermediate representations but also the main direction of future research.

(5) Most of the existing deep learning-based smart contract vulnerability detection methods are black-box detection processes, which give the final vulnerability detection results by training vulnerability detection models. Due to the inherent black-box nature of the deep learning model, its internal specific working status and processing process are opaque, so there is a lack of reasonable explanations for the vulnerability detection results (such as labeling the exact code location or code line where there may be vulnerabilities), which results in the detection result being unconvincing. Therefore, the deep learning model should consider giving a reasonable explanation of its interpretability while outputting the vulnerability detection results. It is worth mentioning that the expert rules defined in traditional detection tools are also powerful tools for analyzing contract vulnerabilities, and future deep learning models should consider integrating the expert rules related to vulnerabilities in traditional detection methods to better improve the accuracy of vulnerability detection.

\section{Discussion and conclusion}
\label{sec:conclusion}
Smart contracts, as one of the most successful applications of the blockchain, have greatly expanded the application scenarios and practical significance of blockchain, playing a vital role in the blockchain ecosystem. With the maturity of blockchain technology and the prevalence of smart contracts, the security and reliability of smart contracts have become an increasingly important hot research topic. This survey summarizes the current common security vulnerabilities in Ethereum smart contracts and restores typical cases in the history of smart contract security. To prevent the occurrence of contract vulnerability, researchers have proposed a series of smart contract vulnerability detection methods, which are concluded into five categories: formal verification, symbolic execution, fuzzing detection, intermediate representation, and deep learning. We introduce and analyze the principles and characteristics of various methods in detail. Furthermore, we compare and evaluate the detectable vulnerability types and performance of representative smart contract automated detection tools.

Technically, the automated vulnerability detection methods are able to deal with the endless smart contract vulnerabilities accurately and efficiently, reducing the false-positive rate and false-negative rate that may be caused by manual verification and analysis. Therefore, it is of great significance to employ a precise and effective smart contract detection method to solve the problem of contract vulnerability mining. This survey analyzes various research methods and points out that although existing efforts have made breakthroughs and encouraging results in the field of smart contract vulnerability detection, current detection methods are not perfect and face the following key problems.
\begin{itemize}
	\item  \textbf{Low accuracy of vulnerability detection.}\quad Currently, most smart contract detection tools still catch a high false-positive rate and false-negative rate. Take the evaluations of smart contract vulnerability detection tools in section~\ref{sec:tools} as an example, VaaS and TMP both achieve high accuracy, which is more than 80\%, while the accuracy of the other three detection tools is only around 60\%, which is far from satisfying the current application scenarios of a large number of contracts and various smart contract vulnerabilities. Therefore, the accuracy of smart contract vulnerability detection tools is a key issue that needs to be improved.
	\item  \textbf{Low coverage of vulnerability type.}\quad Due to the various and complex smart contract vulnerabilities, most detection tools fail to cover all the varieties of vulnerabilities. For example, see the evaluations of section~\ref{sec:tools}, Securify can detect most types of vulnerabilities. However, some other detection tools only support a single vulnerability or verify low-level security issues that lack monitoring and inference during the contract execution so it is difficult to discover and locate cross-contract vulnerabilities. Therefore, making detection tools cover more comprehensive smart contract vulnerabilities is also a crucial challenge.
	\item \textbf{Time-consuming of vulnerability audit.}\quad Efficient and fast audit of smart contract vulnerabilities is also a key element to ensure contract security. Current detection tools have low efficiency in vulnerability mining, which hinders the development and expansion of smart contracts. For example, the average detection time of Mythril is 225.6 seconds, and VaaS is about 159.4 seconds while ContractFuzzer takes about 352.2 seconds. Thus, under the background of the ever-increasing number of smart contracts, ensuring the efficiency of vulnerability auditing is also a difficulty that needs to resolve urgently.
	\item \textbf{Automation level of vulnerability detection.}\quad The characteristic of full automation is also an essential part of ensuring the efficiency of vulnerability detection. It needs to point out that existing detection tools are not fully automated, such as \emph{F}* and KEVM framework. Besides, some automatic detection tools (e.g., Securify and Smartcheck) are unable to clarify whether there are vulnerabilities in the output contract, which require the manual classification for the detected suspected vulnerabilities, increasing the workload in the smart contract detection process and reducing the detection efficiency to a large extent. Therefore, how to achieve a more comprehensive automated detection method need to resolve in future research.
	\item \textbf{Language diversity of smart contracts.}\quad Specifically, there are many kinds of programming languages in the real world. At present, dozens of languages can be used to implement smart contracts, such as Solidity, Go, C, Java, and so on. Different languages, however, have different syntax and semantic rules, which results in different contract structures. How to make smart contract vulnerability detection tools adapt to most programming languages is also challenging and difficult.
\end{itemize}

Although there still exist many difficulties and challenges in the current development of smart contract vulnerability detection, it is also an opportunity to explore novel technology for opening up a new direction. In recent years, researchers have combined deep learning models to detect smart contract vulnerabilities, making encouraging progress. In the following discussions, we look forward to future research directions and propose suggestions by integrating deep learning and smart contract vulnerability detection technology.
\begin{itemize}
	\item \textbf{Constructing a unified and standardized smart contract vulnerability dataset.}\quad First of all, if we want to make a breakthrough in the detection of smart contract vulnerabilities based on deep learning, we rely on the comprehensive dataset of smart contract vulnerabilities. Currently, due to the lack of a standard dataset, existing deep learning-based methods (such as RecChecker, TMP) can only support detecting a few contract vulnerabilities. Therefore, we need to construct a unified and standardized dataset that covers all the vulnerabilities as much as possible, making the deep learning model play a better role, and promoting the research in this field.
	\item \textbf{Building a comprehensive model for both static and dynamic analysis.}\quad As we know, smart contract security vulnerability detection based on deep learning is still in its infancy (such as SaferSC, RecChecker, TMP). Most methods can only analyze the contract at the level of static source code or bytecode. However, such static analysis tends to miss possible existing execution paths. In the meantime, due to the lack of dynamic interaction with external contracts, it usually leads to a high false-positive rate or false-negative rate. Therefore, in order to satisfy the requirements of significant application scenarios, we need to consider combining dynamic and static analysis to build a comprehensive deep learning model.
	\item \textbf{Training a reusable and scalable vulnerability detection model.}\quad With the explosive growth of the number of smart contracts, the corresponding security vulnerabilities are becoming more and more complex and unpredictable. At present, existing vulnerability detection methods based on deep learning are focusing on training the models for the vulnerabilities that have been discovered. Therefore, whether they can quickly adapt to the new kinds of vulnerabilities still needs to be further studied. We believe that the rich security vulnerabilities in the open-source smart contract ecosystem should be fully utilized to build a reusable and scalable vulnerability detection model to cope with the new and endless smart contract vulnerabilities.
\end{itemize}

To summarize, the rapid development of blockchain technology provides a reliable and feasible execution environment for smart contracts. As smart contracts are popularized in various decentralized applications, the security issues of smart contracts are becoming more and more important. This survey sorts out the common smart contract vulnerabilities and compares the \emph{accuracy}, \emph{F1-Score}, and \emph{average detection time} of existing vulnerability detection methods in detail. Furthermore, we give suggestions for the problems that exist in the current research work and discuss the challenges in future research, as well as the possibility of combing the recent achievements in deep learning technology, in order to inspire future research work.

\bibliographystyle{IEEEtran}
\bibliography{conference}

\end{document}